%% file: main.tex
\documentclass{article}

\usepackage{microtype}
\usepackage{longtable}
\usepackage{inconsolata}
\usepackage{url}
\usepackage{booktabs}
\usepackage{amsfonts}
\usepackage{nicefrac}
\usepackage{microtype}      
\usepackage{xcolor}         
\usepackage[ruled]{algorithm2e}
\SetAlFnt{\small}
\usepackage{amsmath}
\usepackage{bm}
\usepackage{bbold}
\usepackage{graphicx}
\usepackage{multirow}
\usepackage{textcomp}
\usepackage{wrapfig}
\usepackage{xspace}
\usepackage{interval}
\usepackage{bm,upgreek}
\usepackage{colortbl}
\usepackage{enumitem}
\usepackage{pifont}
\usepackage{multirow,makecell}
\usepackage{subfigure}
\usepackage{grffile}

\usepackage{hyperref}

\newcommand\metricname{PAM\xspace}
\newcommand\metricnameAvg{PAM$_\mathrm{avg}$\xspace}
\newcommand\metricnameAvgSim{PAM$_\mathrm{avgsim}$\xspace}

\makeatletter
\newcommand*{\@rowstyle}{}
\newcommand*{\rowstyle}[1]{
  \gdef\@rowstyle{#1}%
  \@rowstyle\ignorespaces%
}
\newcolumntype{=}{
  >{\gdef\@rowstyle{}}%
}
\newcolumntype{+}{
  >{\@rowstyle}%
}
\makeatother

\usepackage[accepted]{icml2024}

\usepackage{amsmath}
\usepackage{amssymb}
\usepackage{mathtools}
\usepackage{amsthm}

\usepackage[capitalize,noabbrev]{cleveref}

\theoremstyle{plain}

\theoremstyle{definition}

\theoremstyle{remark}

\usepackage[textsize=tiny]{todonotes}

\icmltitlerunning{\metricname: Prompting Audio-Language Models for Audio Quality Assessment}

\begin{document}

\twocolumn[
\icmltitle{\metricname: Prompting Audio-Language Models for Audio Quality Assessment}



\icmlsetsymbol{equal}{*}

\begin{icmlauthorlist}
\icmlauthor{Soham Deshmukh}{yyy,comp}
\icmlauthor{Dareen Alharthi}{yyy}
\icmlauthor{Benjamin Elizalde}{comp}
\icmlauthor{Hannes Gamper}{comp}
\icmlauthor{Mahmoud Al Ismail}{comp}
\icmlauthor{Rita Singh}{yyy}
\icmlauthor{Bhiksha Raj}{yyy}
\icmlauthor{Huaming Wang}{comp}
\end{icmlauthorlist}

\icmlaffiliation{yyy}{Carnegie Mellon University, Pittsburgh, USA}
\icmlaffiliation{comp}{Microsoft, Redmond, USA}

\icmlcorrespondingauthor{Soham Deshmukh}{sdeshmukh@microsoft.com}

\icmlkeywords{audio quality metric, audio language models, audio-text}

\vskip 0.3in
]



\printAffiliationsAndNotice{}


\input{tex/abstract}

\input{tex/Introduction}
\input{tex/method}

\input{tex/experiments}
\input{tex/results1}
\input{tex/results2}
\input{tex/conclusion}

\nocite{zhou2021narle, 8722433, dhamyal2022describing, deshmukh2023training}

\bibliography{custom}
\bibliographystyle{icml2024}

\newpage
\appendix
\onecolumn
\input{tex/appendix}

\end{document}

%% file: tex/abstract.tex
\begin{abstract}{
While audio quality is a key performance metric for various audio processing tasks, including generative modeling, its objective measurement remains a challenge.
Audio-Language Models (ALMs) are pre-trained on audio-text pairs 
that may contain information about audio quality, the presence of artifacts or noise. 
Given an audio input and a text prompt related to quality, an ALM can be used to calculate a similarity score between the two.
Here, we exploit this capability and introduce \metricname, a no-reference metric for assessing audio quality for different audio processing tasks. 
Contrary to other ``reference-free'' metrics, \metricname does not require computing embeddings 
on a reference dataset nor training a task-specific model on a costly set of human listening scores. We extensively evaluate the reliability of \metricname against established metrics and human listening scores on four tasks: text-to-audio (TTA), text-to-music generation (TTM), text-to-speech (TTS), and deep noise suppression (DNS).
We perform multiple ablation studies with controlled distortions, in-the-wild setups, and prompt choices. Our evaluation shows that \metricname correlates well with existing metrics and human listening scores. These results demonstrate the potential of ALMs for computing a general-purpose audio quality metric}. 

\end{abstract}

%% file: tex/introduction.tex
\section{Introduction}

Audio Quality Assessment (AQA) refers to the subjective assessment of the perceived overall quality of a signal \cite{torcoli2021objective}. The gold standard of AQA consists of assessment by humans, which is a challenging task that requires many listening tests in controlled setups. Moreover, these experiments are time-intensive and costly, and hence cannot be carried out multiple times for every setup or result.
Hence, measurements that can closely estimate human assessment of audio quality are essential for the development and evaluation of models that perform audio generation tasks.

\begin{figure}[!h]
\vspace{-2pt}
\centering
\includegraphics[height=0.15\textheight,width=0.45\textwidth]{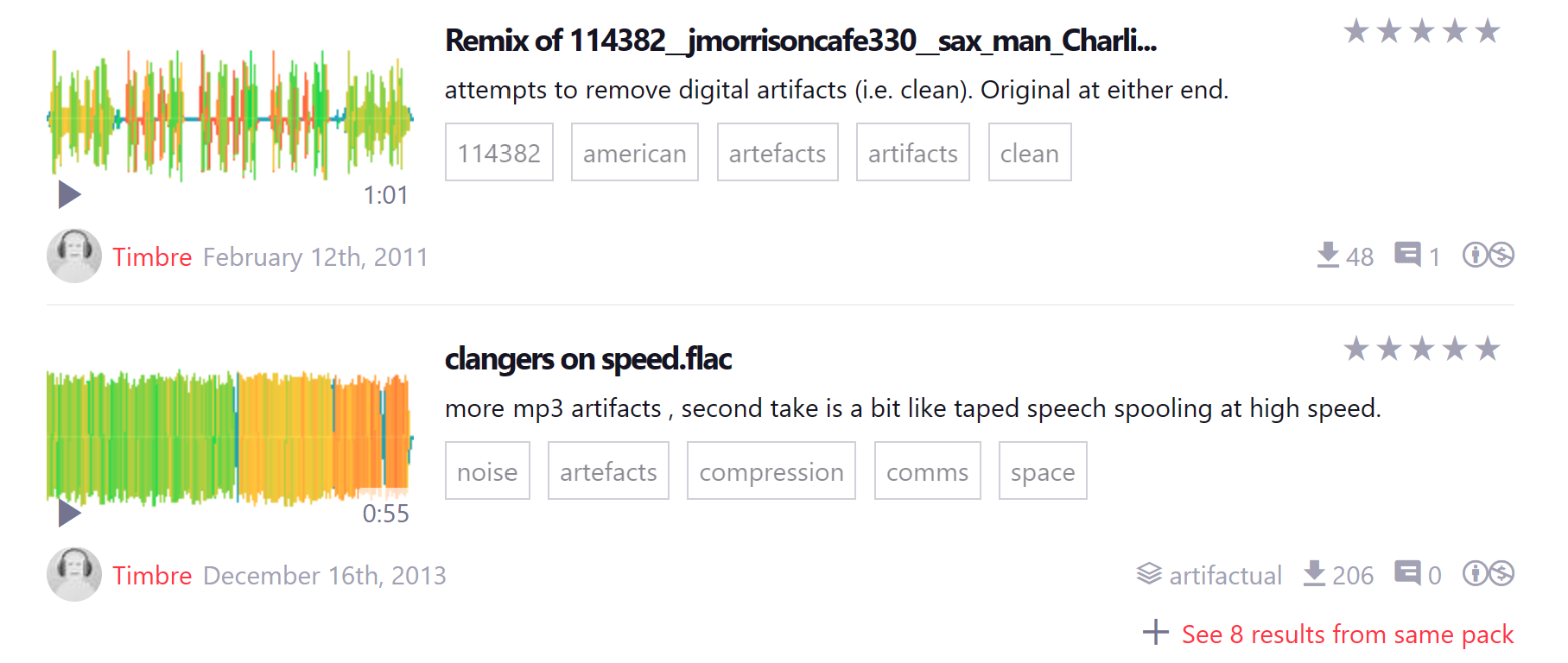}
\caption{Search result of ``Artifacts" on FreeSound.org. These audio-text pairs are included in ALM training.\vspace{-10pt}}
\label{fig:intro}
\end{figure}

Audio generation tasks entail sounds, music, and speech. All tasks employed different audio quality metrics, including some that aim to resemble human assesments. TTA uses metrics like FD and Fr\'echet Audio Distance (FAD)~\cite{kilgour2018fr}, IS, KL, and subjective metrics like Overall Quality (OVL) and Relation of audio to text caption (REL)  \cite{kreuk2022audiogen}. TTM uses FAD and subjective metrics like MCC \cite{musicgen}.  TTS uses metrics like WER, SpeechLMScore \cite{maiti2023speechlmscore}, and perceptual metrics like MOSNet \cite{lo2019mosnet}, FSD \cite{le2023voicebox}, and MCC. However, several aspects of audio quality are shared across tasks, such as the presence of artifacts. Ideally, one metric should measure quality regardless of the task hence, addressing the challenges of task-specific metrics.

\begin{figure*}[!t] 
\vspace{-15pt}
\centering
\includegraphics[height=0.23\textheight,width=0.9\textwidth]{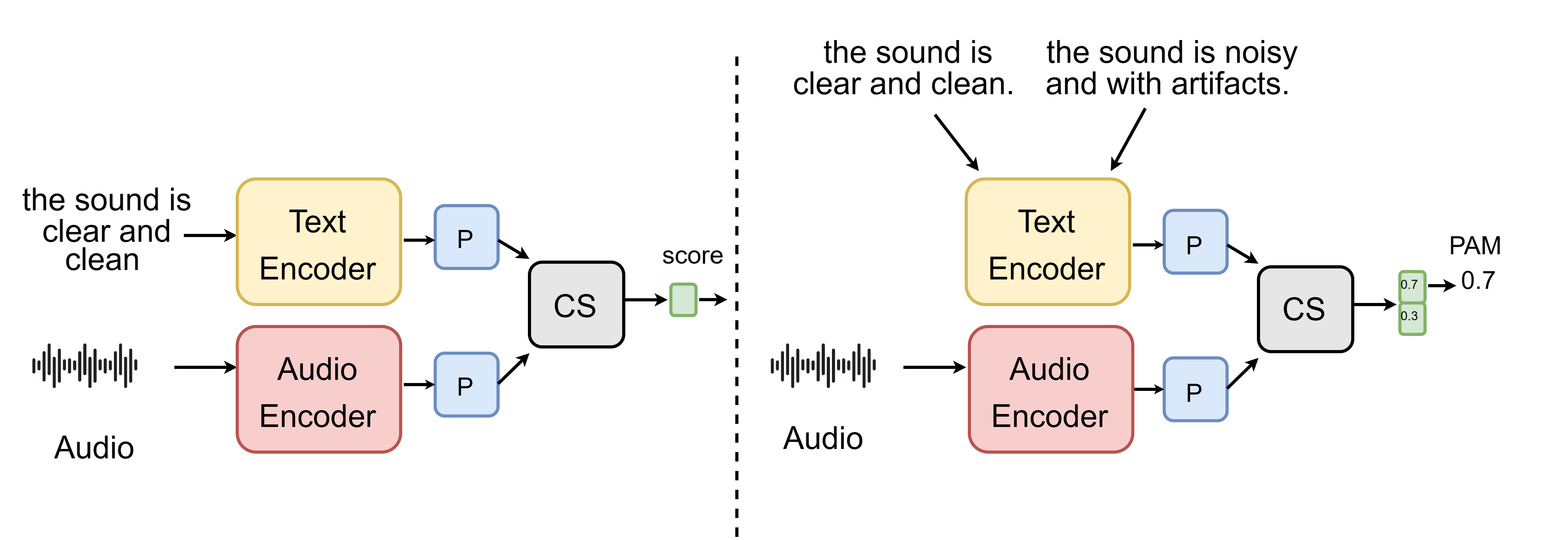}
\caption{Two prompt strategies leveraging ALM to perform AQA. The figure on the left shows a naive approach that takes as input one prompt about quality and the audio intended for assessment. The output is the cosine similarity between the audio and text embeddings, which determines the correspondence between them. The figure on the right shows \metricname computation, which uses two ``opposing" prompts to derive a score.}
\vspace{-15pt}
\label{fig:diagram}
\end{figure*}


Current metrics provide a reliable evaluation but pose different challenges. Reference-based metrics require ground truth for computation. To assess the quality of a recording, the generated audio is compared against a desired recording to measure how much the quality degraded. Reference-free metrics do not require a desired recording, but usually require a pretrained model to compute embeddings on a reference dataset. The selection of the model and the dataset would highly affect the score~\cite{gui2023adapting}. 
Other metrics like DPAM \cite{manocha2020differentiable}, MOSNet \cite{lo2019mosnet}, and DNSMOS \cite{reddy2021dnsmos} train a model using human evaluation and at inference use the model predictions as a proxy for human evaluation. This requires the curation of human evaluation and model training for each audio task. 

Instead, we propose a no-reference metric that leverages learning perceptual audio quality from human assessments in text descriptions. ALM have learned from millions of audio-text pairs sourced from the Internet. Some of the audio has a corresponding natural language description of quality (See Fig. \ref{fig:intro}). For example, audio-text models \cite{elizalde2022clap, Elizalde2023NaturalLS, wu2022large, pengi} trained on FreeSound data, 
have seen text descriptions like ``Pad sound, with a lo-fi, high compression type feel to it. The noise floor, with a low pass filter set around 50Hz and several octaves of pitch bend". Although the ALM is not explicitly trained for audio quality assessment, it has ingested hundreds of human annotations describing their perception of the audio. Because ALM can be used out of the box in a Zero-Shot fashion, they can compare text prompts about quality against audio without requiring a reference. 


In this work, we introduce PAM, a novel, reference-free method for assessing audio quality, which provides an advancement in audio quality evaluation. Our contribution includes (1) The first Audio Language Model (ALM) based metric for general-purpose audio-quality assessment which is truly reference-free (2) A two-prompt antonym strategy that makes using ALM feasible and improves correlation with human perception (3) Extensively testing PAM on four audio tasks: Text-to-Audio Generation, Text-to-Music Generation, Text-to-Speech, and Deep Noise Suppression. (4) For each task, we collected human listening scores and will publicly release the evaluation framework, generated audio and human listening scores. We hope our work and data release push the development of general-purpose audio quality assessment\footnote{Repository: \url{https://github.com/soham97/PAM}}.


%% file: tex/method.tex
\section{\metricname}
\vspace{-0.1in}
Our proposed metric \metricname can perform audio quality assessment by exploiting the joint multimodal space learned by an ALM. The learned space can be used to quantify the correspondence between quality-related text prompts and audio recordings. 

\subsection{Audio Quality Assessment} \label{sec: audio quality}

\textbf{Audio quality.} The term implies a variety of properties in various contexts. For this work, we consider audio to be high quality when the presence of artifacts and noise is imperceptible. For example, white noise, clipping, and other distortions. We did not consider non-speech audio as noise, such as sound events, music, reverb, echo, and in general all naturally occurring sounds. 

\noindent \textbf{Learning quality from audio-text pairs.} ALM are pretrained with millions of audio and their corresponding natural language descriptions. The text is usually metadata created by the user who uploads the audio file to a web archive and some pairs describe the quality and the presence of artifacts and noise in a given audio. 
Therefore, as a first step, this work focuses on specific prompting strategies to show the potential of audio-text learning for audio quality assessment. 

\begin{figure*}[!ht]
    \centering
    \subfigure{\includegraphics[height=0.11\textheight,width=0.19\textwidth]{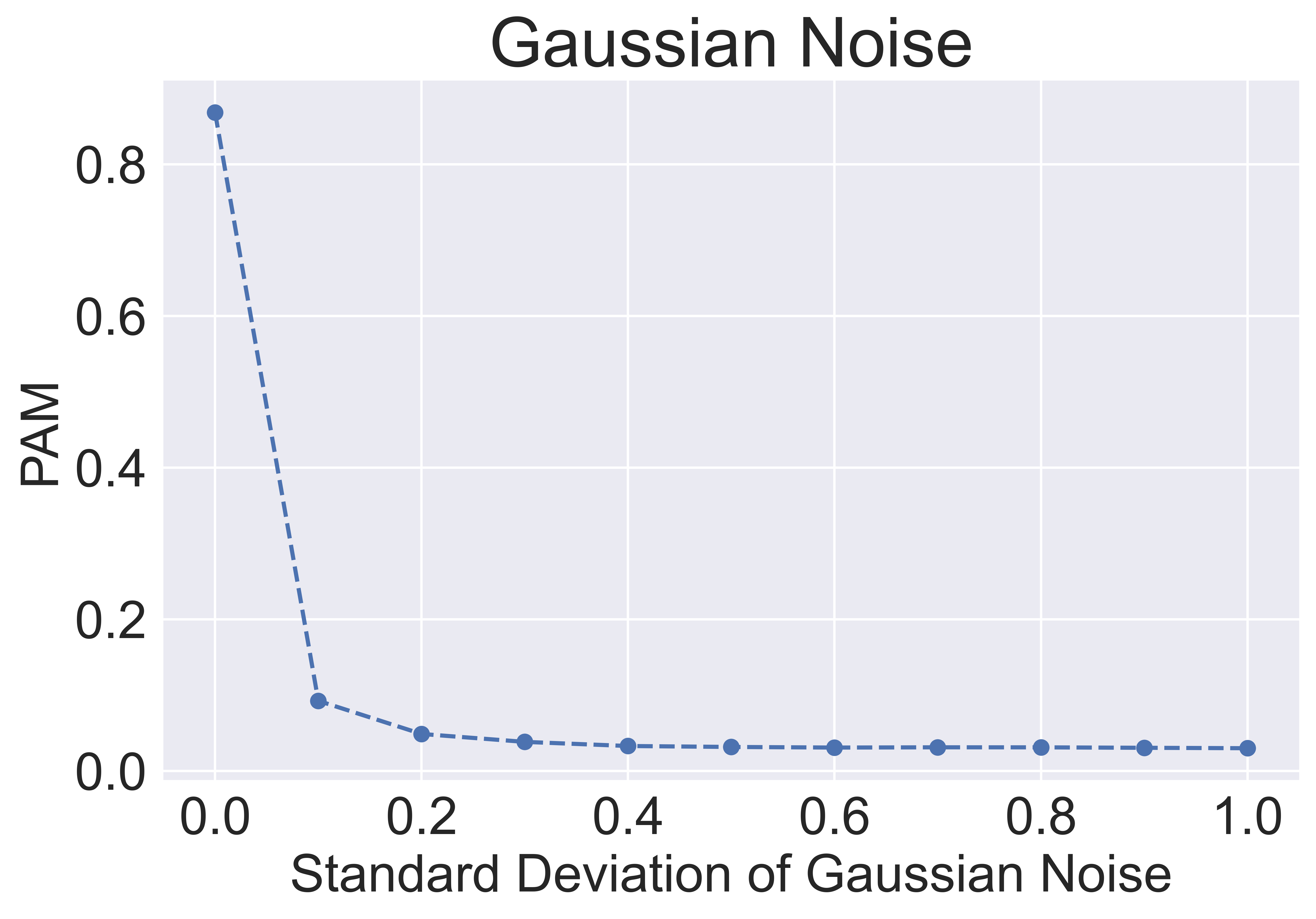}} 
    \subfigure{\includegraphics[height=0.11\textheight,width=0.19\textwidth]{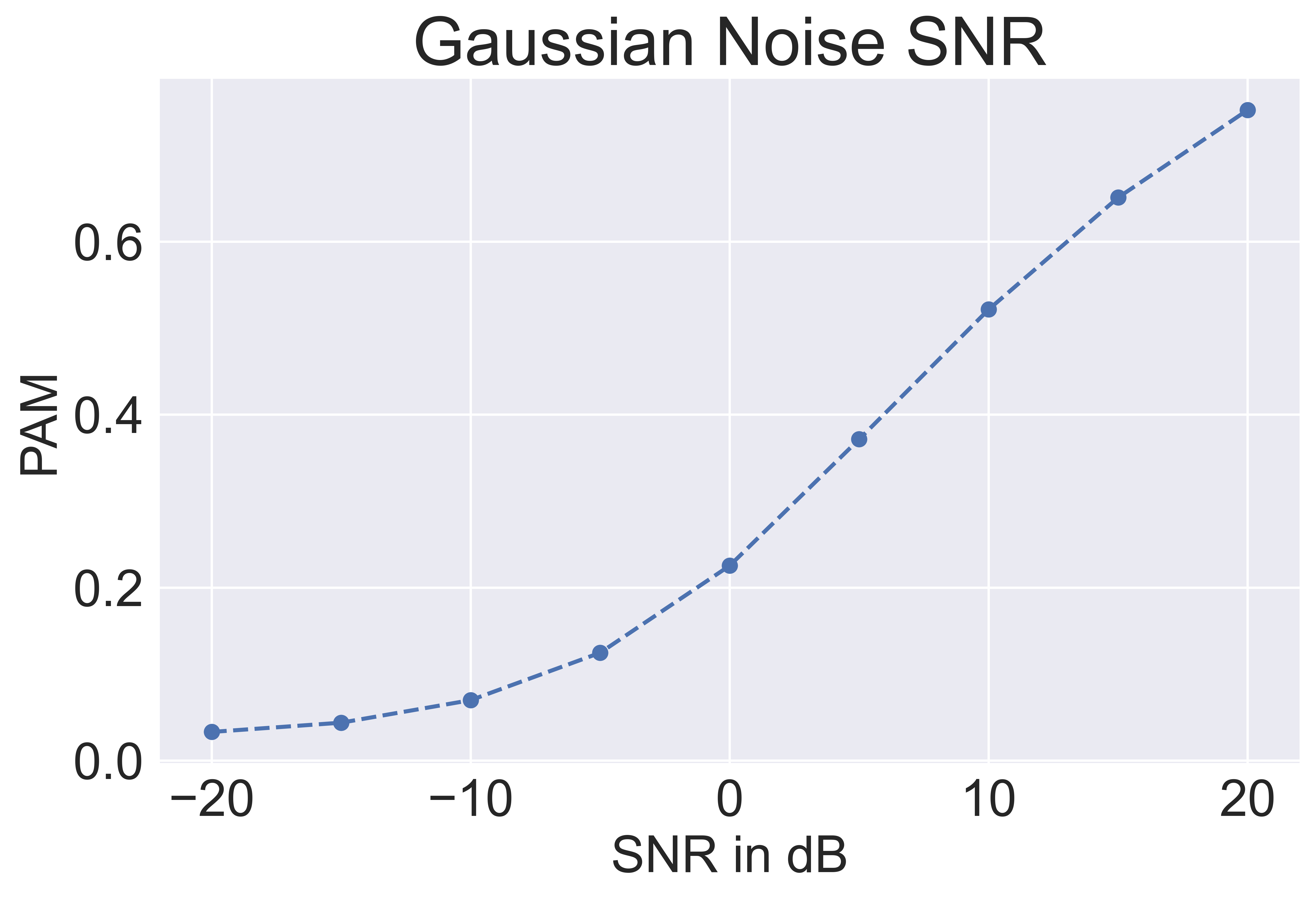}} 
    \subfigure{\includegraphics[height=0.11\textheight,width=0.19\textwidth]{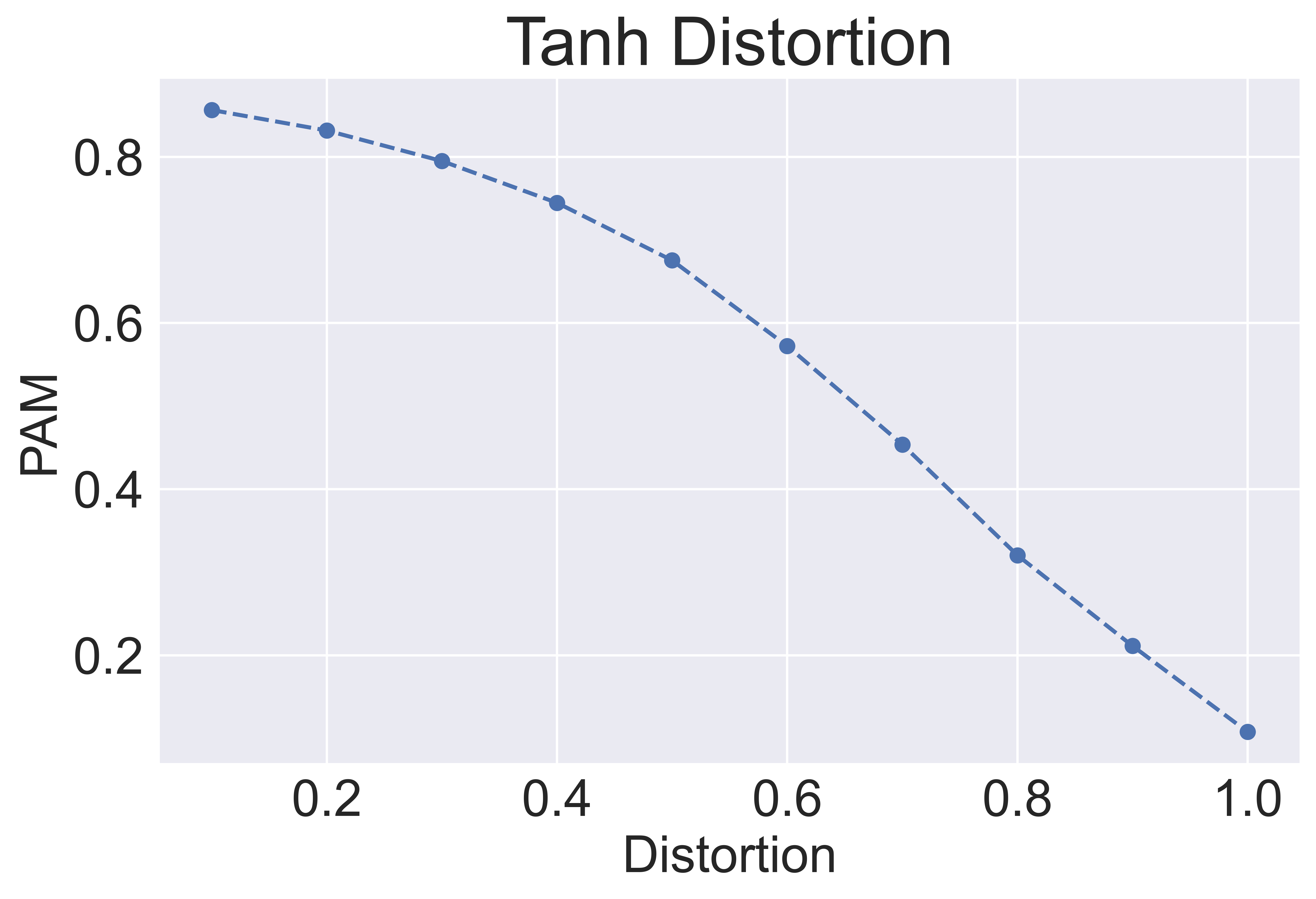}}
    \subfigure
    {\includegraphics[height=0.11\textheight,width=0.19\textwidth]{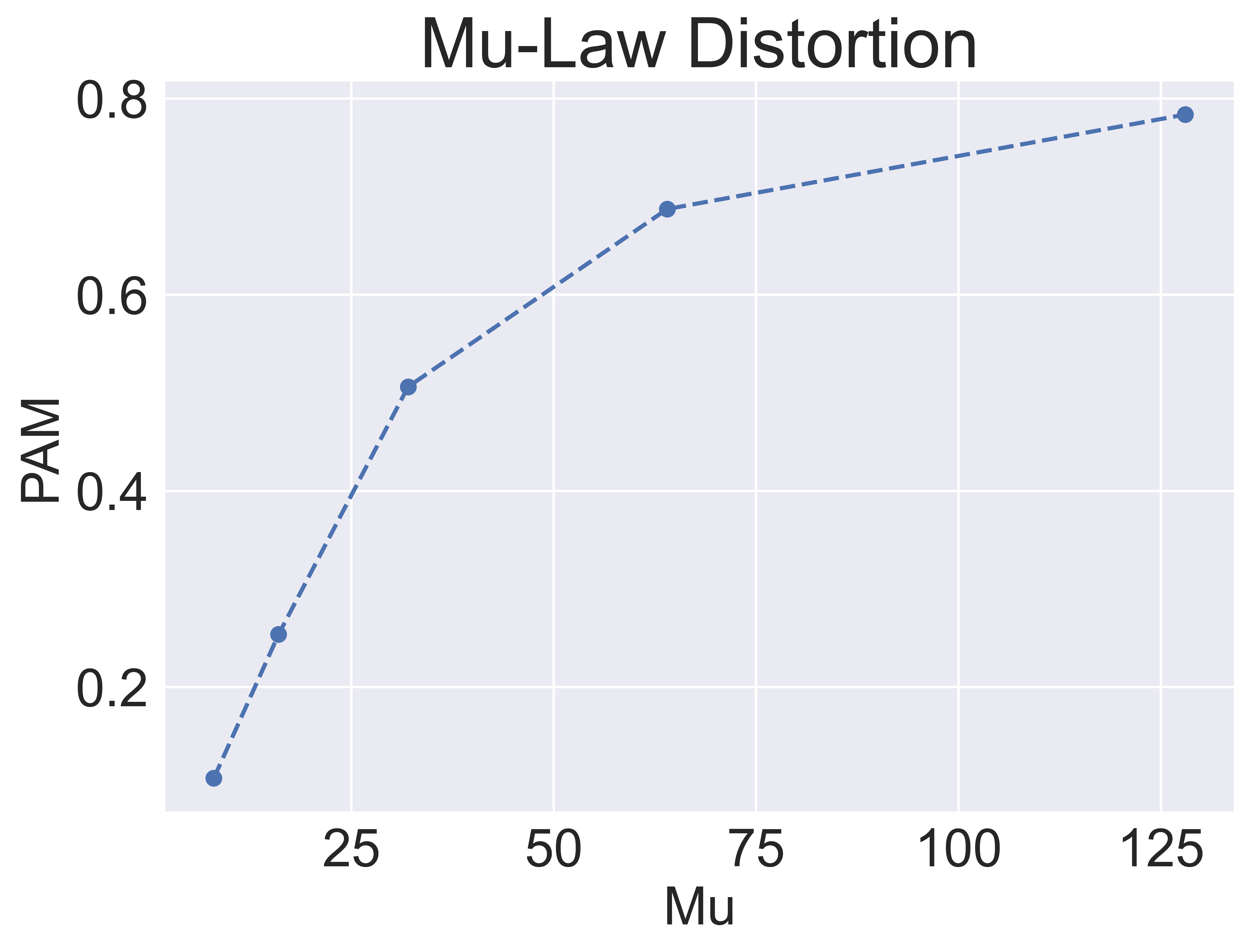}}
    \subfigure{\includegraphics[height=0.11\textheight,width=0.19\textwidth]{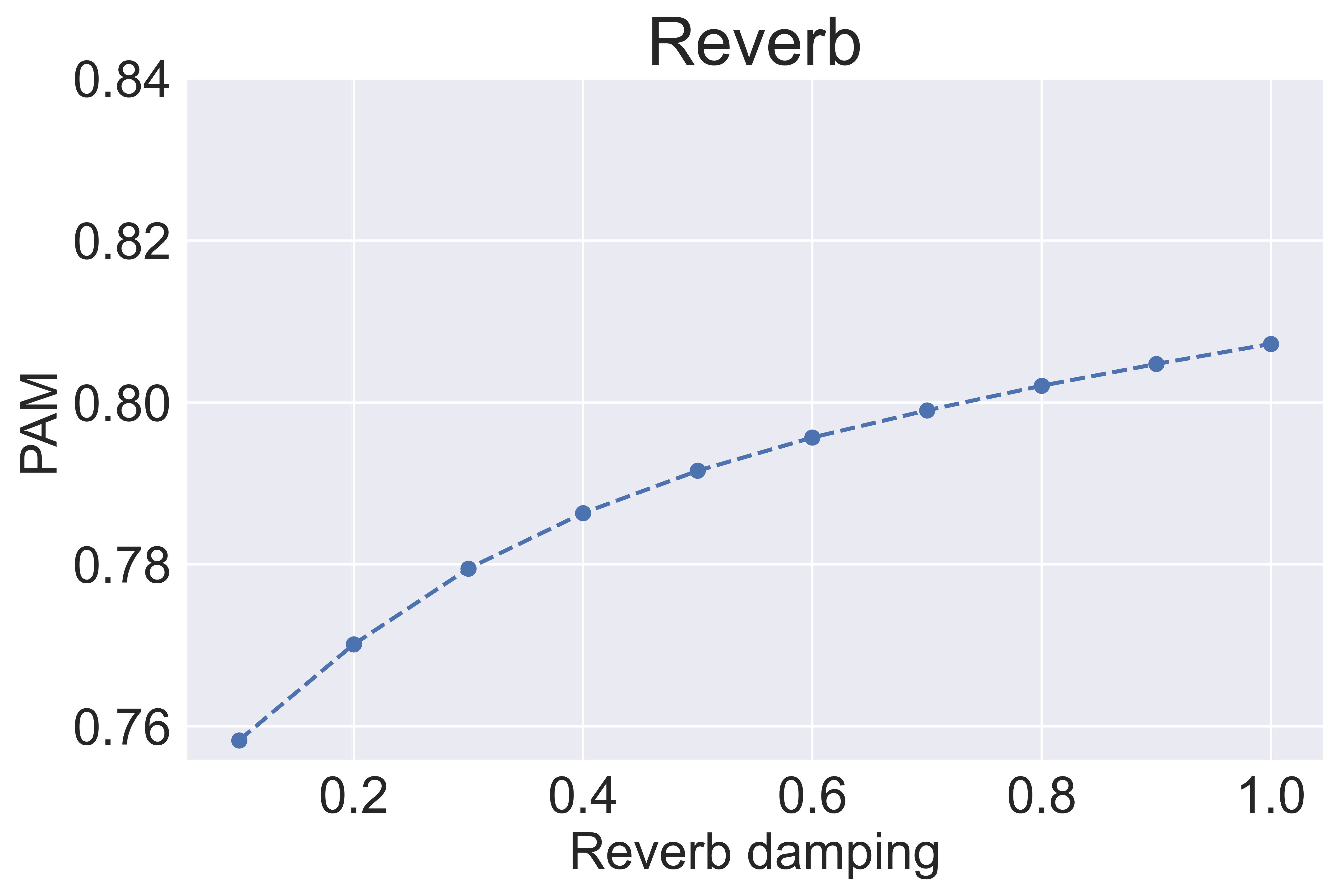}}
    \vspace{-0.15in}
    \caption{Effects of \metricname when adding different distortions: (a) Gaussian Noise (b) Gaussian Noise with SNR (c) Tanh distortion (d) Mu-Law compression (e) Reverb. The dataset is a professionally recorded dataset containing sounds. \metricname decreases as the distortion of the signal increases. \vspace{-0.2in}
    }
    \label{fig:distortions_main}
\end{figure*}

\noindent\textbf{Audio-Language Model.}
In this work, we used the ALM called CLAP \cite{Elizalde2023NaturalLS} trained on 4.6M pairs. The pairs are sourced from different publicly available datasets including web archives, such as FreeSound and FindSound which have descriptions about audio quality (See Fig. \ref{fig:intro}). CLAP consists of audio and text encoders pretrained using Contrastive Learning and it can be used for Zero-Shot inference. That is, at inference time, the user provides an audio file for assessment and text prompts about the quality (e.g. ``the sound is clear and clean"). The model embeds the audio and text in a multimodal space using the respective encoders, computes the cosine similarity between the embeddings and produces a correspondence score. Determining the prompting strategy and setup to use CLAP for audio quality assessment is still an open question. 

\subsection{Prompting setup} \label{sec: prompting setup}
The user can provide an audio file and text class of ``the sound is clear and clean" and determine the audio-text similarly using the model. The similarity can be squashed between 0 and 1 and used as a score. Though this method is valid and used for multiple tasks \cite{liu2023audioldm, kreuk2022audiogen,ghosh2023compa}, we see prompting with just one class of ``the sound is clear and clean" leads to a poor correlation with human perception and distortions across various tasks and distributions. One of the reasons this strategy does not work is due to linguistic ambiguity. Particularly, if the prompt is “the sound is clear and clean”, then depending on the context, the model can infer: (1) The sound is easy to understand, see, or hear, without any distortion, noise, or interference. (2) The sound is pure, crisp, and pleasant, without any harshness, dullness, or muddiness. This meaning is based on the definition of ‘clean’ as ‘having a pure, fresh, or smooth quality.’, or (3) The sound is honest, accurate, and truthful, without any deception, manipulation, or bias. This meaning is based on the definition of ‘clean’ as ‘free from dishonesty or corruption’.

To address this problem, we prompting a strategy that will minimize ambiguity in the latent space. This is achieved by using multiple prompts that force the model to make similarity calculations along the latent subspace of audio quality (e.g., “the sound is clear and clean” and “the sound is noisy and with artifacts”). This not only reduces the ambiguity but allows us to audio quality measurement as a binary classification problem, where the final score is between 0 to 1 and regarded as a relative similarity. The \metricname computation is explained in Section \ref{sec: computePAM}


\subsection{Choice of quality prompts}
The choice of text prompts has an impact on the similarity and if not optimal leads to spurious correlates being measured rather than audio quality. The \metricname score uses `opposite" text prompts of “the sound is clear and clean” and “the sound is noisy and with artifacts”. These prompts are chosen based on analysis of CLAP's \cite{Elizalde2023NaturalLS} training data and tested across various setups, tasks, and distributions. Specifically, the text descriptions in training data are combined and a frequency count per word is obtained (unigram) after removing stop-words. Then, we computed BERT embeddings for each word and took cosine similarity between filtered words and the word “noise”. We use the top 10 words from the list to form the two text prompts. In the rest of the paper, \metricname implies the usage of the above prompts. 

However, to get more insight into the type of artifacts and noise, the prompts can be changed. That is the prompts can be designed for specific tasks and setups in mind. For example, in the definition of audio quality \ref{sec: audio quality}, reverb and echo are not considered as noise and \metricname score does not degrade with Reverb addition \ref{fig:distortions_main}. Therefore, our general \metricname score cannot be used as a metric for the specific task of Acoustic Echo Cancellation (AEC) to measure echo suppression. Therefore, we can design attribute-specific prompts for audio quality outside of our definition. 


\subsection{Computing \metricname} \label{sec: computePAM}

The \metricname computation is shown in Fig~\ref{fig:diagram}, right section. The user provides an audio file which is converted into a mel-spectrogram ($x \in \mathbb{R}^{T \times F}$) and passed to CLAP's audio encoder to produce an audio embedding $v \in \mathbb{R}^{1 \times d}$. In parallel, the two ``opposing" prompts about quality (“the sound is clear and clean” and “the sound is noisy and with artifacts”) are tokenized and embedded using the text encoder to produce text embeddings $u \in \mathbb{R}^{d \times N}$. After projection into the multimodal space, the dot product is computed between the two embeddings, followed by softmax: 
$p_\mathrm{h} = \frac{e^{z_\mathrm{h}}}{\sum_{j=1}^2 e^{z_j}}$,
where $\mathrm{h}$ is the index of the prompt related to high quality, $z_j = u_j \cdot v$, $(\cdot)$ denotes the dot product, and $p \in \mathbb{R}^{1\times 2}$. The value of $p_\mathrm{h} \in [0,1]$ is the \metricname score and informs about the quality of the audio. 

\begin{figure*}[!ht]
\centering
\includegraphics[height=0.35\textheight, width=0.8\textwidth]{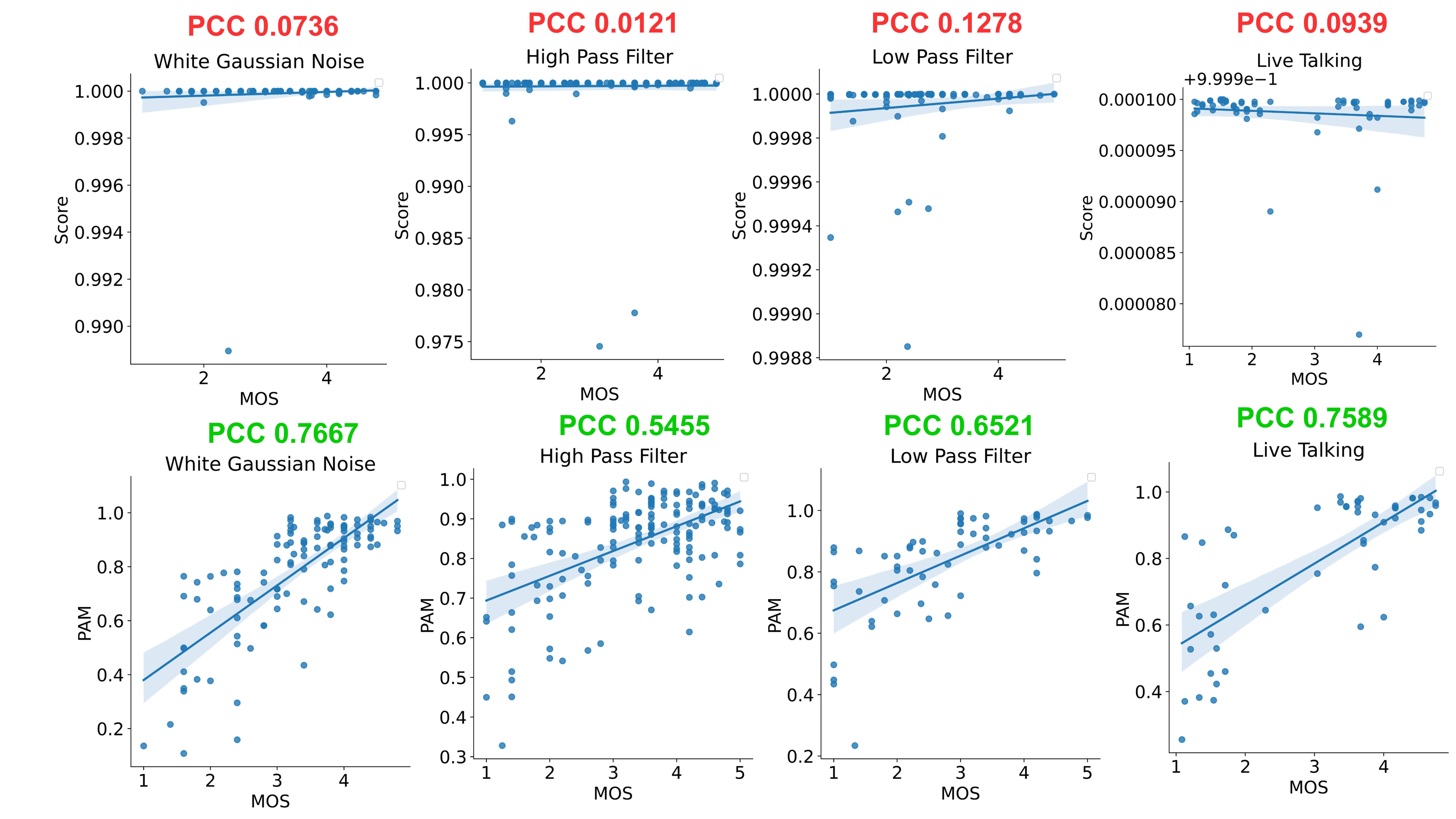} 
\caption{
(PCC) Correlation plots between MOS (human subjective evaluations) and two prompt strategies for four distortions applied to the NISQA dataset. The top row refers to using the naive single-prompt strategy and the second row shows \metricname (two opposite prompt strategy). Both described in Fig. \ref{fig:distortions_main}. Using one prompt for AQA does not correlate with MOS, but using two prompts (\metricname) does. \vspace{-0.1in}
}
\label{fig:human correlation distortion}
\end{figure*}

%% file: tex/experiments.tex
\section{Experiments} \vspace{-0.1in}
\label{sec:exp and results}

The experimental setup is designed to provide a comprehensive evaluation of \metricname across different distortions, prompting strategies, and datasets from different audio generation tasks. All experiments are run using a single 16GB V100 GPU. \\ 
\textbf{Distortions.} We systematically add various types of distortions: Gaussian Noise, Gaussian Noise with Signal-to-Noise Ratio (SNR), Tanh distortion, Mu-Law compression, and Reverb across various source distributions and check its effect on the \metricname. The results are in Section \ref{sec:result: effect of distortions}. \\
\textbf{Prompting strategy.} \metricname uses a two opposite prompt strategy with the text of ``the sound is clear and clean" and ``the sound is noisy and with artifacts”. In section \ref{sec: results: prompting strategy}, we compare it against the naive single-prompt strategy. We also compare it against human evaluation.  \\
\textbf{Audio tasks.} In Section \ref{sec: results: audio tasks}, we consider different generation tasks like Text-to-Audio, Text-to-Music and Text-to-Speech Generation. For each task, we use multiple models, perform human listening tests, and compare \metricname against established metrics. 

%% file: tex/results1.tex
\section{Results}

\subsection{Effect of distortions} \label{sec:result: effect of distortions}
An audio quality metric should degrade with the presence of distortions and artifacts in the audio. To verify this, we add common simulated distortions to audio sourced from a professionally recorded sound effect pack. The four types of distortions used are (1) Gaussian Noise with increasing standard deviation (2) Gaussian Noise addition with particular SNR (3) Tanh distortion (4) Mu Law compression. Lastly, we add Reverb, which by the definition in Section \ref{sec: audio quality} is not considered an artifact or distortion. Figure \ref{fig:distortions_main} shows the effect of distortions on \metricname score when tested on a professionally recorded sound effect pack. The PAM score degrades as the noise is added except for Reverb. For Reverb, the PAM score is fairly constant, i.e., changes from 0.76 to 0.81. While for others we see considerable degradation in \metricname score. To check robustness across source distribution, we change the dataset from professionally recorded to AudioCaps (audio from YouTube videos containing sound events), MusiCaps (music tracks from YouTube), and LibriTTS (speech, audioboks). We see similar trends of PAM score degrading with the addition of distortions and consistent scores across Reverb. The details can be found in Appendix \ref{appendix: effect of distortions}.

\begin{table*}[ht]
\center
\begin{tabular}{=l+c+c+c+c+c+c+c+c} \hline
\makecell{Model} & Dur. (h) & Param & FD $\downarrow$ & FAD $\downarrow$ & IS $\uparrow$ & KL sig $\downarrow$ & KL soft $\downarrow$ & \metricname $\uparrow$ \\ \hline
AudioLDM-l \cite{liu2023audioldm} & 9031 & 975M & 43.83 & 6.229 & 5.067 & 7.422 & 2.723 & 0.2417 \\
AudioLDM2-l \cite{audioldm2} & 29510 & 1.5B & 50.07 & 3.477 & 5.195 & 6.379 & 2.200 & 0.4267 \\
MelDiffusion (Appendix \ref{appendix: text-to-audio models arch}) & 145 & 383M & 20.27 & 3.296 & 8.460 & 3.579 & \textbf{1.390} & \textbf{0.5412} \\ 
AudioGen-m \cite{kreuk2022audiogen} & 6824 & 1.5B & \textbf{18.67} & \textbf{2.850} & \textbf{9.202} & \textbf{3.391} & 1.797 & 0.4683 \\
\rowstyle{\color{gray}} AudioCaps \cite{audiocaps} & - & - & 00.00 & 0.000 & 9.488 & 0.000 & 0.000 & 0.6772 \\ \hline
\end{tabular}
\caption{\label{table: T2A results}\footnotesize Evaluation of Text-to-Audio generation models from the literature. Established metrics and \metricname show similar trends. 
}
\vspace{-0.1in}
\end{table*}

\begin{figure*}[!ht]
    \centering
    \subfigure{\includegraphics[height=0.14\textheight,width=0.49\textwidth]{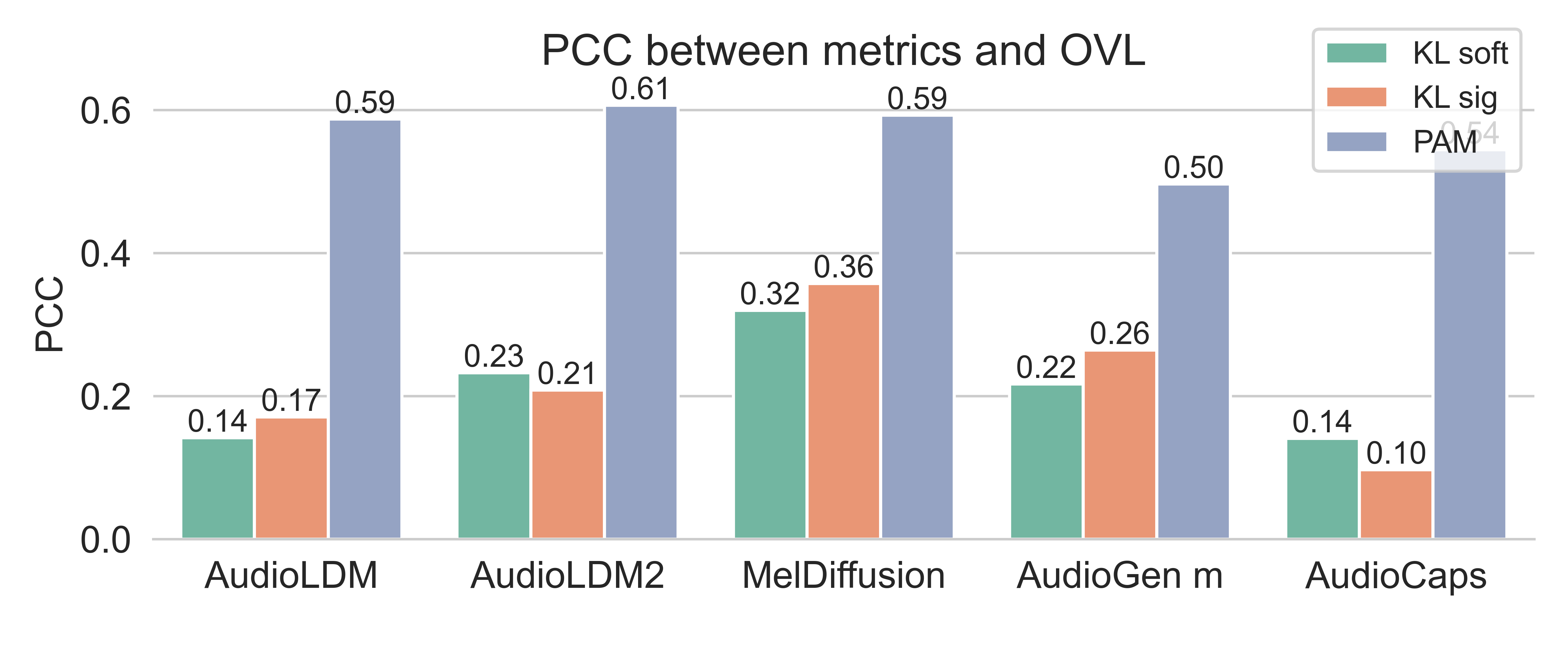}} 
    \subfigure{\includegraphics[height=0.14\textheight,width=0.49\textwidth]{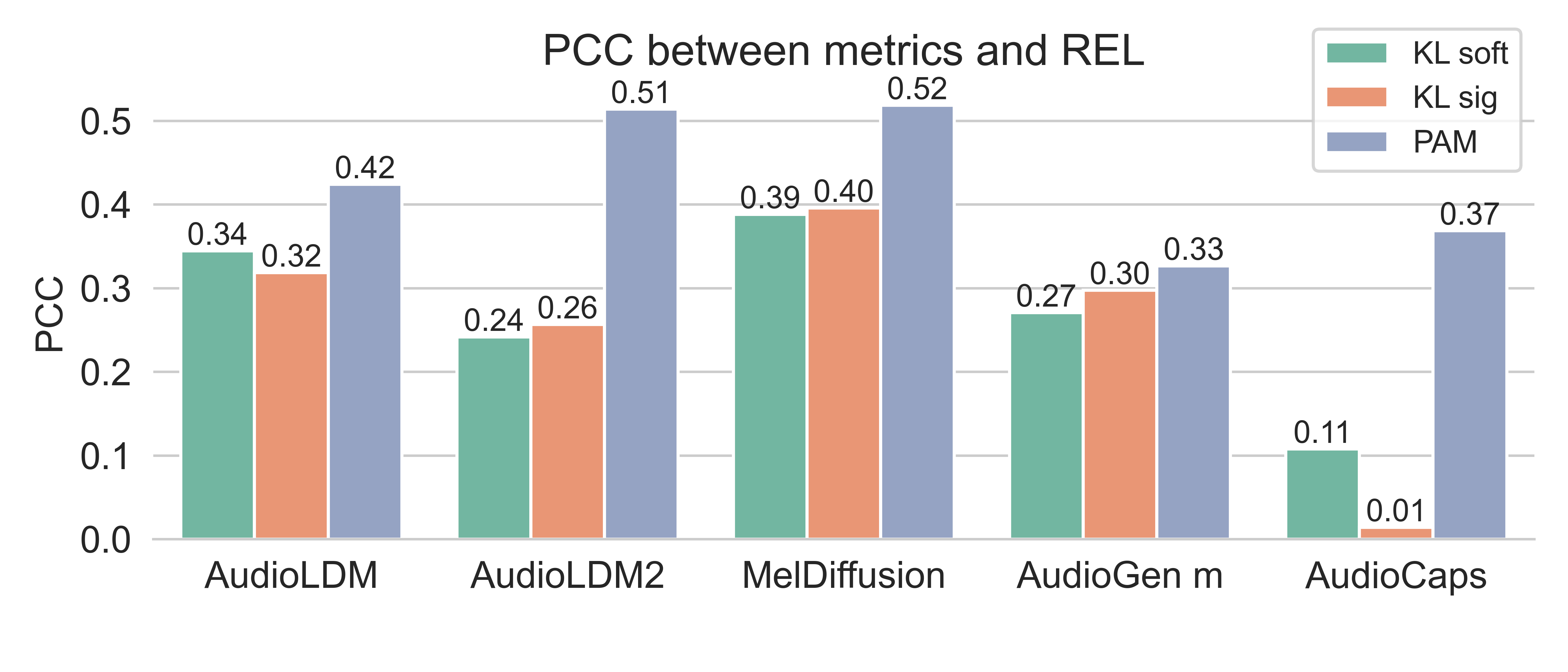}}
    \vspace{-0.15in}
    \caption{The absolute value of (PCC) correlation between OVL, REL and different TTA generation models. The input text captions come from the AudioCaps dataset. \metricname, as a single-metric, is capable of correlating well with both, overall quality (OVL) and relevance to the input caption (REL).
    \vspace{-0.2in}}
    \label{fig:PCC T2A}
\end{figure*}

\subsection{Prompting strategy} \label{sec: results: prompting strategy}

Figure \ref{fig:diagram} shows two different prompting strategies that can be used to get a quality-related score. The figure on the left shows naive prompting and the figure on the right shows the opposite prompting strategy of \metricname. The advantages of the opposite prompting setup and the limitations of the naive prompt are explained in Section \ref{sec: prompting setup}. In this section, we perform experiments to compare two setups with human listening scores.

We use the NISQA (Non-Intrusive Speech Quality and TTS Naturalness Assessment) \cite{nisqa} dataset to check the correlation between \metricname, the single prompt strategy, and human perceptual evaluation. The NISQA Corpus includes more than 14,000 speech samples with simulated (e.g. codecs, packet-loss, background noise) and live (e.g. mobile phone, Zoom, Skype, WhatsApp) conditions. Each file is labeled with subjective ratings of the overall quality. We use simulated and live talk corpus from NISQA. The simulated corpus contains simulated distortions with speech samples from four different datasets and the live talk corpus contains recordings of real phone and VoIP calls. Unlike \metricname, NISQA considers sounds events as noise, so human raters labelled the recordings as low quality. Therefore, we created a filtered NISQA set and applied four distortions: (1) white noise addition with a particular SNR (2) live talk on a laptop or smartphone (3) low bandpass filter (4) high bandpass filter. We check the correlation of the single prompt strategy and the opposite prompt strategy against the Mean Opinion Score (MOS) from human listeners. MOS is a numerical measure of the human-judged overall quality and it is the arithmetic mean of the ratings given by subjects on a predefined scale. We used the existing MOS numbers from NISQA. The Pearson Correlation Coefficient (PCC) measures linear correlation between two sets of data \cite{pearson1920notes} and it is shown in Figure \ref{fig:human correlation distortion}. PCC ranges from -1 to 1, where -1 indicates a perfect negative correlation, 0 indicates no correlation, and 1 indicates a perfect positive correlation. We see that the single-prompt strategy does not correlate with MOS, the human perceptual evaluation. While \metricname not only correlates, but achieves a PCC greater than 0.7 on (1) white noise distortion and (2) real-world talk recorded from laptops and smartphones.

\subsection{Assessing quality across distributions}
An audio quality metric should give high scores to audio that is free from distortions. For example, professionally recorded and edited audio should achieve a higher \metricname score compared to audio sourced from YouTube, which is generally recorded with handheld devices and may contain noise or distortions. To confirm this hypothesis we carried on the following setup. We compare \metricname among three sets in Table \ref{table: source table}. (1) AudioCaps dataset sourced from YouTube containing sound events (clapping, alarms, dog barking, etc). (2) MusicCaps data sourced from YouTube with additional filtering to retain high-quality and remove low-quality music recordings. (3) Professionally recorded audio containing sound events.

\begin{table}[!ht]
\center
\begin{tabular}{=l+c+c} \hline
\makecell{Dataset} & Source & \metricname $\uparrow$ \\ \hline
AudioCaps (test set) & YouTube & 0.6772 \\
MusicCaps (test set) & YouTube-filtered & 0.7718 \\
Professionally recorded & Studio & \textbf{0.8684} \\ \hline
\end{tabular}
\caption{\label{table: source table}\footnotesize \metricname score is higher for professionally recorded audio than for audio from YouTube videos.
} \vspace{-0.2in}
\end{table}


%% file: tex/results2.tex
\vspace{-0.1in}
\section{\metricname for audio tasks} \label{sec: results: audio tasks}
In this section, we use PAM to evaluate models for generation tasks.
For each task, we compare \metricname against task-specific metrics and human evaluation to show its reliability as an AQA metric. 

\begin{table*}[ht]
\center
\begin{tabular}{=l+c+c+c+c+c+c+c+c} \hline
\makecell{Model} & Dur. (h) & Param & FD $\downarrow$ & FAD $\downarrow$ & IS $\uparrow$ & KL sig $\downarrow$ & KL soft $\downarrow$ & \metricname $\uparrow$ \\ \hline
AudioLDM2-m \cite{audioldm2} & - & 1.1B & 37.54 & 6.706 & 1.841 & 4.456 & 1.611 & 0.6157 \\
MusicLDM \cite{musicldm2023} & 466 & - & 31.05 & 6.109 & 1.840 & 4.333 & 1.428 & 0.6887 \\ 
MusicGen-l \cite{musicgen} & 20000 & 1.5B & 25.91 & 4.878 & 2.101 & 4.389 & \textbf{1.281} & \textbf{0.8492} \\
MusicGen-mel. \cite{musicgen} & 20000 & 1.5B & \textbf{24.65} & \textbf{3.955} & \textbf{2.242} & \textbf{4.197} & 1.339 & 0.7704\\
\rowstyle{\color{gray}} MusicCaps & - & - & 00.00 & 0.000 & 4.547 & 0.000 & 0.000 & 0.7718 \\ \hline
\end{tabular}
\caption{\label{table: T2M results}\footnotesize Evaluation of Text-to-Music generation models from the literature. Established metrics and \metricname show similar trends.
}
\vspace{-0.1in}
\end{table*}

\subsection{Text-to-Audio generation}
TTA generation models synthesize non-speech non-music audio (sounds) from text descriptions. Although there are established metrics available, evaluating the generation quality of these models is still an open research question.

Table \ref{table: T2A results} shows the evaluation of TTA with objective metrics from in literature \cite{liu2023audioldm, kreuk2022audiogen}.
These metrics do not consider any type of perceptual aspect and consist of a distance between the generated audio and a distribution from a reference set. The objective metrics for all the systems are in Appendix \ref{appendix: text-to-audio generation}. We use publicly available variants of AudioLDM \cite{liu2023audioldm}, AudioLDM2 \cite{audioldm2}, AudioGen \cite{kreuk2022audiogen} and MelDiffusion (See Appendix for details \ref{appendix: text-to-audio generation}). We choose the variant of the model corresponding to the largest parameter count, because it usually correlates better with higher performance. 
The captions from the AudioCaps test set (747 captions) are used to generate audio from the above 4 models and their variants. Captions are textual descriptions of the sounds, i.e. ``A drone is whirring followed by a crashing sound". 

We carry out a human listening experiment to compute the correlation between metrics and human perception. We randomly picked 100 captions and their corresponding generated audio from the test set. During the experiment, each participant was asked to rate each audio in terms of Overall Quality - OVL and Relation of audio to text caption - REL on a five-point Likert scale. The order of audios was fully randomized and each audio was rated by 10 participants. Raters were recruited using the Amazon Mechanical Turk platform. To ensure quality in the annotations, participants who consistently provided identical scores in every HIIT (e.g., all 1s) or who completed the task in less than 10 seconds were excluded. 

Figure \ref{fig:PCC T2A} summarizes the PCC between per-model metrics and OVL and REL respectively. \metricname correlates correlates significantly better with human perception of quality (OVL and REL) than the task-specific metrics of KL softmax and KL sigmoid. 
The KL metric uses the CNN14 \cite{pann} model to extract audio embeddings for the generated and reference set. The CNN14 model is trained to classify audio into different sound events and hence does well at recognizing the presence of sound events rather than overall quality. Also, a recent work \cite{audioldm2} observed that reference-free metrics like KL provide high scores when the generation model is trained on the same distribution data as the KL reference set.  
\metricname is a no-reference metric so it does not have these drawbacks.



\subsection{Text-to-Music generation}
TTM generation models synthesize music based on text descriptions. Although objective performance metrics exist, evaluating the subjective quality of these models remains an open research question.

Subjective performance can be described in terms of Acoustic Quality (AQ), which measures whether the generated sound is free of noise and artifacts, and Musical Quality (MQ), which measures the quality of the musical composition and performance. 

A commonly used reference set for evaluating TTM models is MusicCaps~\cite{musiclm}, a music subset of AudioSet~\cite{audioset} that contains rich text captions provided by musical experts. A recent work~\cite{gui2023adapting} used GPT-4 to derive AQ and MQ ratings for MusicCaps audio samples via text-analysis of the corresponding captions. Each MusicCaps song was assigned one AQ and one MQ label "high", "medium", or "low". If AQ or MQ could not be inferred from the caption text, the label "not mentioned" was assigned. These text-derived labels were shown to correlate reasonably well with human perception~\cite{gui2023adapting}.
We compare these text-based AQ and MQ labels with an audio-only analysis via \metricname. The results shown in Table \ref{table: AQ for music} indicate similar trends of audio-only \metricname and text-based analysis via GPT-4.


\begin{table}[!ht]
\center
\begin{tabular}{=l+c+c+c} \hline
\makecell{\% of data} & GPT-4 inference & \metricname $\uparrow$ \\ \hline
39\% & Low acoustic quality & 0.7294 \\
2\% & Medium acoustic quality & 0.8138 \\
1\% & High acoustic quality & \textbf{0.8333} \\
\rowstyle{\color{gray}} 58\% & Unknown acoustic quality & 0.7975 \\ \hline
9\% & Low musical quality & 0.7222 \\
9\% & Medium musical quality & 0.7770 \\
42\% & High musical quality & \textbf{0.7932} \\
\rowstyle{\color{gray}} 41\% & Unknown musical quality & 0.7593 \\ \hline
\end{tabular}
\caption{\label{table: AQ for music}\footnotesize Acoustic and musical quality of MusicCaps derived from
text-analysis based GPT-4 labels and audio-based \metricname scores exhibit similar trends.
\vspace{-0.1in} 
}
\end{table}

For a direct comparison with human perception, we calculate \metricname on a set of real and generated music recordings. Subjective AQ and MQ labels were collected by authors in~\cite{gui2023adapting} as MOS scores from several human judges. The real samples were taken from the Free Music Archive (FMA) and MusicCaps. For TTM generation, publicly available variants of MusicLM \cite{musiclm} and MusicGen \cite{musicgen}, as well as Mubert~\cite{Mubert} were used. 
\begin{figure}[!ht]
    \centering
    \includegraphics[height=0.18\textheight, width=0.48\textwidth]{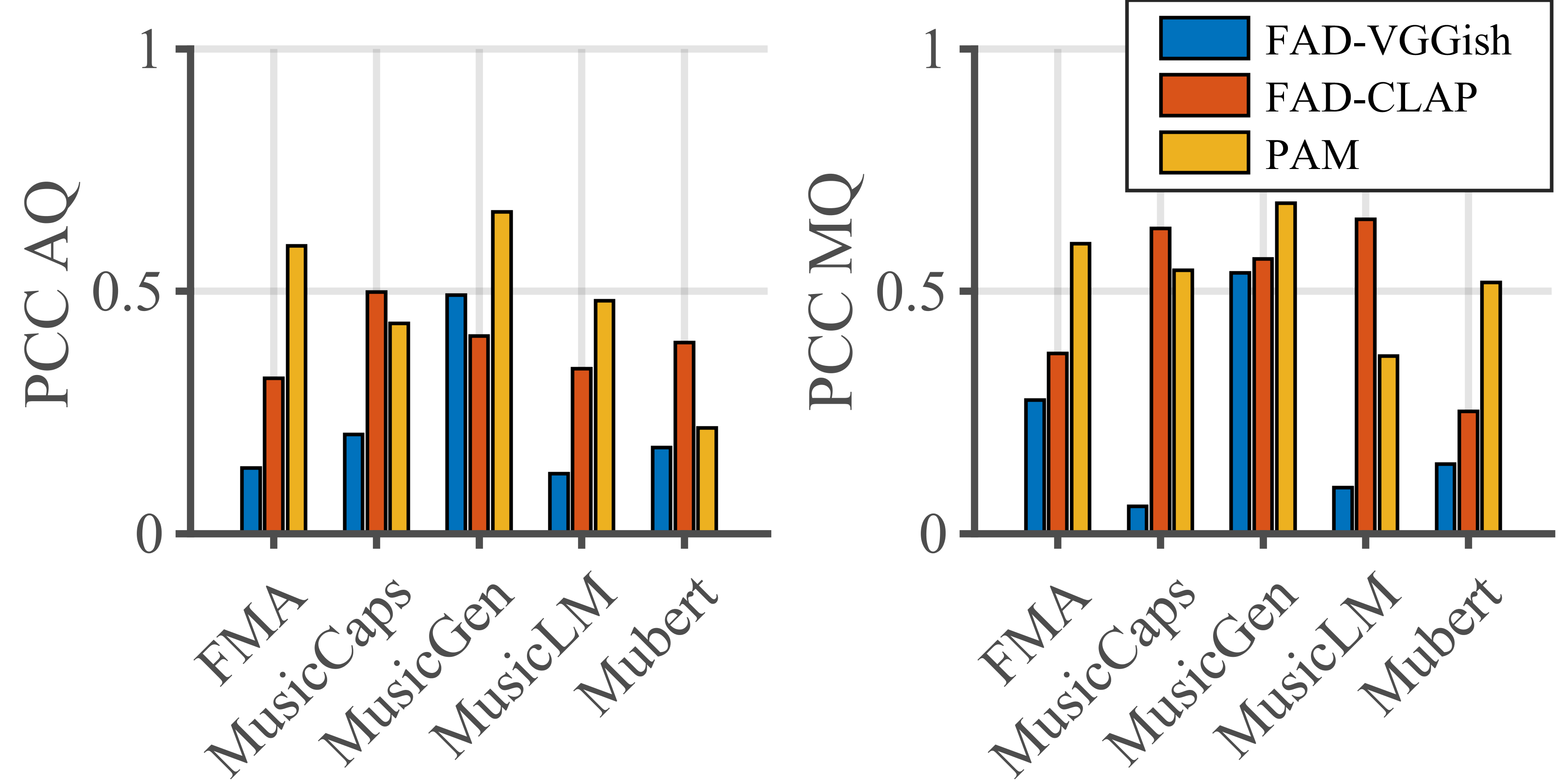}
    \caption{PCC between listening test MOS and Fr\'echet Audio Distance (FAD) and \metricname, for acoustic (AQ, left) and musical quality (MQ, right). 
    \vspace{-0.1in}
    }
    \label{fig:PAM_vs_FAD}
\end{figure}

Figure~\ref{fig:PAM_vs_FAD} shows PCC between the MOS ratings and \metricname. For comparison, the (absolute) PCC of the Fr\'echet Audio Distance (FAD) is shown for two pretrained models. Recall that FAD requires a pretrained model to compute audio embeddings on a reference dataset. Here, we used MusCC, a set of studio quality music~\cite{gui2023adapting}, as a reference for FAD. \metricname performs competitively, outperforming the commonly used FAD-VGGish metric in all comparisons.  

Table \ref{table: T2M results} shows the evaluation of TTM with objective metrics and the proposed \metricname. 
The samples are generated using MusicCaps captions as prompts. The row in grey shows results for the original MusicCaps audio and constitutes an upper performance bound for objective metrics that use MusicCaps audio as a reference, including FD, FAD, and KL. 
However, as the quality of MusicCaps samples varies significantly (cf.\ Table~\ref{table: AQ for music}) TTM models may outperform MusicCaps in perceptual quality, as \metricname indicates for MusicGen-l. We observe similar trends for PCC results (Table \ref{table: t2m new}) corresponding to Table \ref{table: T2M results}. 

\begin{table}[!ht]
\center
\begin{tabular}{=l+c+c+c+c+c} \hline
\makecell{Model} & \makecell{Subj.} & KL sof $\downarrow$ & KL sig $\downarrow$ & PAM $\uparrow$ \\ \hline

AudioLDM2 & OVL & 0.1241 & 0.0929 & \textbf{0.3295} \\
MusicLDM & OVL & 0.1291 & 0.1617 & \textbf{0.4235} \\ 
MusicGen-l & OVL & 0.3353 & 0.3275 & \textbf{0.6018} \\
MusicGen-mel & OVL & 0.0993 & 0.1423 & \textbf{0.6908} \\
MusicCaps & OVL & 0.2314 & 0.2319 & \textbf{0.5549} \\ \hline
AudioLDM2 & REL & 0.005 & \textbf{0.1416} & 0.1238 \\
MusicLDM & REL & 0.0662 & 0.0398 & \textbf{0.1399} \\ 
MusicGen-l & REL & 0.2237 & 0.2034 & \textbf{0.2309} \\
MusicGen-mel & REL & 0.1831 & 0.2562 & \textbf{0.2622} \\
MusicCaps & REL & 0.1035 & 0.1566 & \textbf{0.3284} \\ \hline
\end{tabular}
\caption{\label{table: t2m new}\footnotesize PCC between human evaluation MOS and different metrics for the models in Table. The subjective metric (Subj.) indicates the metric used for PCC computation. \ref{table: T2M results} \vspace{-0.15in}}
\end{table}

\subsection{Speech Synthesis}

\begin{figure}[!ht]
\centering
\includegraphics[height=0.15\textheight, width=0.48\textwidth]
{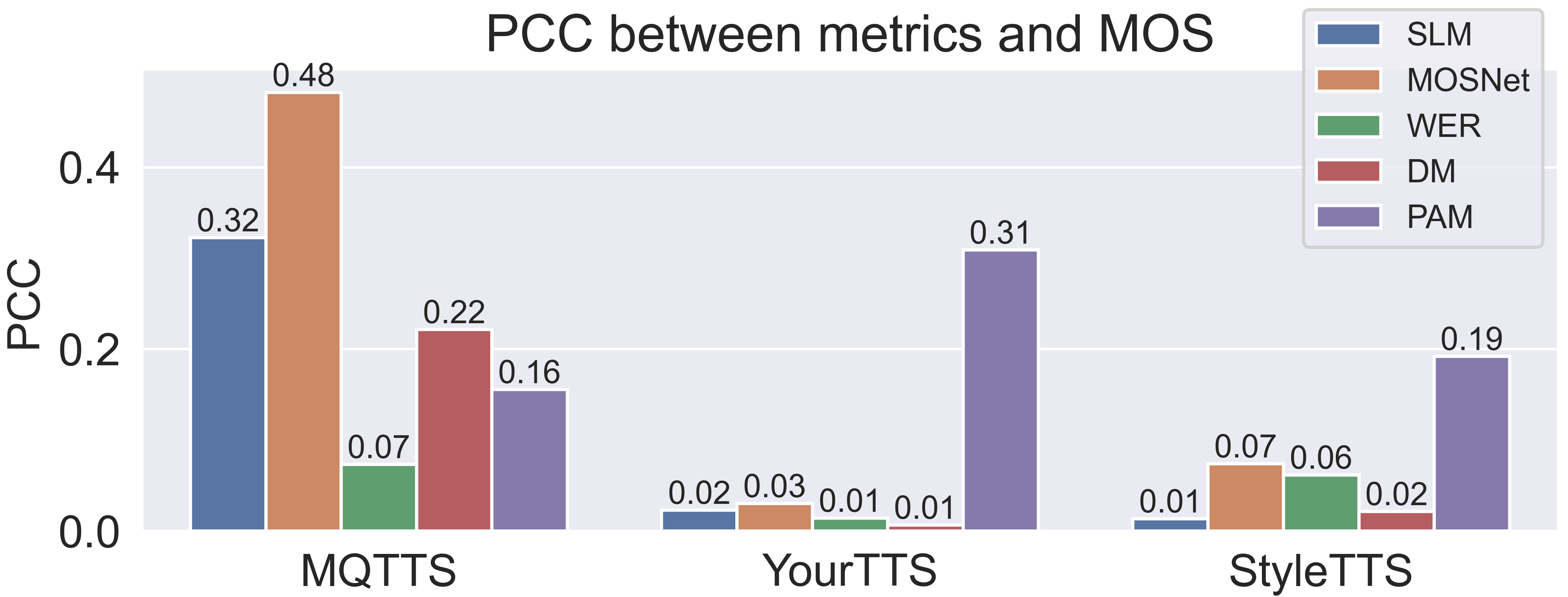} 
\caption{Absolute PCC between the human evaluation and metrics for TTS models. The transcripts are sourced from the LibriTTS dataset. On average, \metricname correlates better with human perception of speech quality than existing metrics. \vspace{-0.1in}}
\label{fig:aug}
\end{figure}

Speech Synthesis involves creating artificial speech, either by converting text to speech (TTS) or altering existing speech to sound like a different speaker or style, known as voice conversion. In our study, we examine the effectiveness of \metricname in the above two tasks. For TTS, A recent work \cite{alharthi2023evaluating} conducted human evaluation studies for different TTS systems. The study used StyleTTS \cite{li2022styletts}, MQTTS \cite{mqtts}, and YourTTS \cite{casanova2022yourtts} to generate speech for 100 sentences from the LibriTTS dataset \cite{zen2019libritts}. Each generated sample was rated by 10 raters. We use this dataset and compare \metricname with existing metrics. The absolute results are shown in Table \ref{table: zero-shot TTS} and the PCC correlation with human evaluation in Figure \ref{fig:aug}. On average, \metricname correlates better with human perception of speech quality than existing metrics. 

\begin{table}[!ht]
\center
\begin{tabular}{=l+c+c+c+c} \hline
\makecell{Metric} & StyleTTS & MQTTS& YourTTS\\ \hline
WER $\downarrow$ & 18.7 & 29.35 & 22.1\\
SLMScore $\uparrow$  & 3.62 & 4.13 &  3.96 \\
MOSNet $\uparrow$ & 4.49  & 3.57&  4.01 \\
DM $\downarrow$ & 3.30  &  3.90 &  4.50\\ 
\metricname $\uparrow$  & 0.90 & 0.87 & 0.81\\ \hline
\rowstyle{\color{gray}} MOS-N $\uparrow$& 3.68 & 3.66 & 3.59\\ \hline
\end{tabular}
\caption{\label{table: zero-shot TTS}\footnotesize Evaluating different TTS models using metrics from the literature. MOS-N indicates MOS scores for the naturalness of generated speech. \vspace{-0.15in}}
\end{table}

The second speech synthesis task we consider is Voice Conversion (VC), where the aim is to convert audio containing the original speech to audio containing the target speaker's voice. For this, we use the VoiceMOS 2022 challenge dataset \cite{huang2022voicemos}, specifically the VCC subset. The VCC subset includes 3,002 utterances from 79 systems. We test \metricname on this dataset and compare it with existing metrics of MOSNet \cite{lo2019mosnet}, MOS-SSL \cite{cooper2022generalization}, and SpeechLMScore \cite{maiti2023speechlmscore}. \metricname performs worse than other speech-based finetuned metrics. 

\begin{table}[!ht]
\centering
\begin{tabular}{=l+c+c+c+c+c+c}\hline
Source & Model & \multicolumn{2}{c}{Utterance-level} & \multicolumn{2}{c}{System-level} \\ 
&& PCC  & SRCC    & PCC   & SRCC \\ \hline
VCC  &  MOSNet & 0.654 & 0.639& 0.817& 0.796 \\
VCC & MOS-SSL   & 0.891 & 0.883 & 0.983 & 0.964 \\
VCC &SLMS.   &  0.505  & 0.501 & 0.863 & 0.829 \\
VCC & \metricname & 0.389 &0.411 & 0.563 & 0.593\\ \hline
OOD & MOSNet & 0.259 & 0.153 & 0.537 &   0.430\\
OOD & MOS-SSL & 0.467 & 0.459 & 0.357 & 0.437 \\
 OOD & SLMS. & 0.138 & 0.224 &  0.049 & 0.199 \\
OOD & \metricname & \textbf{0.582}& \textbf{0.585} & \textbf{0.634} & \textbf{0.703} \\\hline
\end{tabular}
\caption{Utterance-level and system-level correlation of different metrics with MOS scores. The dataset used is VCC subset and OOD subset from VoiceMOS. \metricname correlates better with MOS than other metrics in the Out-of-Domain setup, suggesting better generalization. \vspace{-0.2in}} 
\label{tab:voicemos}
\end{table}

On both the setup of TTS and Voice Conversion, the literature metrics like MOSNet, and MOS-SSL were trained on the train split of data. Therefore, all the evaluation setup is in-distribution for the existing metrics. To check out-of-distribution performance, we consider an out-of-domain (OOD) subset of the VoiceMOS challenge. the OOD subset is sourced from the 2019 Blizzard Challenge \cite{wu2019blizzard}, and contains 136 Chinese TTS samples. The PCC results of metrics are shown in Table \ref{tab:voicemos}. In the OOD setup, \metricname correlates better than existing metrics that are not trained on the OOD data. This showcases the ability of \metricname to be a zero-shot audio quality metric.  

Overall, \metricname can detect audio quality and distortions in generated speech. For speech tasks, it falls short of task-specific metric, where the generated speech is rated based on intelligibility or speaker characteristics. 
This is explained in the Section \ref{sec: limitations}. 


\subsection{Noise suppression}
\vspace{-0.1in}
\label{sec:DNS}
Noise and artifacts negatively impact perceived speech quality, e.g., in voice communication systems~\cite{DNSchallenge2}. Deep Noise Suppression (DNS) aims at enhancing speech quality by suppressing noise. MOS derived from listeners judging the output of a DNS model provides a subjective performance metric to develop or tune the model. Machine-Learning based blind MOS estimators such as DNS-MOS have shown to outperform existing objective metrics for estimating the speech quality of DNS models~\cite{MOSnonintrusive,reddy2021dnsmos}. We compute \metricname on the output of models participating in the ICASSP 2021 DNS challenge~\cite{DNSchallenge2} and compare it against a state-of-the-art DNS-MOS estimator~\cite{reddy2021dnsmos}. Unlike generative models, DNS involves \emph{removing} unwanted signals, hence its perceptual quality is impacted both by the quality of the (desired) speech as well as the quality and suppression of noise. We hypothesise that estimating such multifaceted quality may benefit from more  comprehensive prompts. As a proof of concept, we calculate \metricname with two prompt averaging strategies (see appendix for details). 
Table~\ref{table: zero-shot SEC} summarizes the results in terms of Spearman’s Rank Correlation Coefficient (SRCC) between the average human-labeled MOS of each tested DNS model and the average DNS-MOS or \metricname. The SRCC indicates how well the ranking of the tested DNS models in terms of their subjective quality is preserved~\cite{reddy2021dnsmos}. \metricname performs competitively compared to the state-of-the-art MOS estimator trained specifically for this task. 
\begin{table}
\setlength{\tabcolsep}{3pt}
\center
\begin{tabular}{=l+c+c+c+c} \hline
\makecell{} & DNS-MOS $\uparrow$ & PAM $\uparrow$ & \metricnameAvgSim  $\uparrow$ & \metricnameAvg $\uparrow$ \\ \hline
SRCC & 0.9753 & 0.8785 & 0.8962 & 0.9289 \\
\hline
\end{tabular}
\caption{\label{table: zero-shot SEC}\footnotesize
SRCC of DNS models participating in the ICASSP 2021 DNS challenge for state-of-the-art DNS MOS estimation model and \metricname. \metricnameAvgSim and \metricnameAvg use alternative prompting strategies (see appendix). \vspace{-0.2in}}
\end{table}

%% file: tex/conclusion.tex

\vspace{-0.1in}
\section{Limitations} \label{sec: limitations}
\vspace{-0.1in}
\metricname show correlation with human perception of audio quality. For the task of Text-to-Audio generation and Text-to-Music generation, \metricname has better PCC with human perception than existing metrics. However, \metricname has limitations.\\
\textbf{Speech generation.} For speech generation tasks like Text-to-Speech and Voice conversion, the PCC is lower than existing objective perceptual metrics trained for the specific task. One reason for the low correlation is that the base model CLAP \cite{Elizalde2023NaturalLS} is not explicitly trained on speech-text pairs, let alone multilingual speech. This limits the capability of \metricname for speech generation tasks. But it shows an opportunity area for further adding such training pairs to CLAP or other ALM. \\
\textbf{Fine-grained qualities} This work focuses on analyzing a specific prompt (``the sound is clear and clean", ``the sound is noisy and contains artifacts") and contrastive prompting setup for audio quality score across audio tasks. However, for specific audio tasks, changing the prompt might lead to better performance. For example, in the TTM task, specific prompts about melody, genre, and tune can provide information about specific qualities other than artifacts.\\
\textbf{Finetuning ALMs.} Existing literature trains a model to predict MOS scores and then later uses the pretrained model as an objective perceptual metric. In our work, we use the base in a zero-shot, and exploring finetuning or few-shot learning has the potential to improve \metricname's performance. 

\vspace{-0.1in}
\section{Conclusion} \label{sec: conclusion} 
\vspace{-0.1in}
This paper proposes \metricname, a reference-free metric for assessing audio quality for any-to-\{audio\} generation. The metric is zero-shot and does not require task reference embeddings or task-specific finetuning to predict human scores. We extensively evaluate \metricname across various distortions and various tasks like text-to-audio, text-to-music, noise suppression, text-to-speech, and voice-conversion. We conduct human listening experiments for each task and check the correlation of \metricname with human perception of audio quality. Against existing metrics, \metricname correlates better with human perception for the audio and music tasks and performs comparably for speech tasks. To further advance the exploration of audio quality metrics, we will release audio and human listening scores.

%% file: tex/appendix.tex
The appendix is organized as follows: Section \ref{appendix sec: related work} covers related work and Section \ref{appendix: effect of distortions} explores effect of distortions, Section \ref{appendix: text-to-audio generation} provides details of TTA generation models and listening experiment for the task, Section \ref{appendix: text-to-music generation} covers TTM generation models and listening experiment for the task, Section \ref{appendix: noise suppression} explains noise suppression task and prompt averaging strategy.  

\section{Related work} \label{appendix sec: related work}

\textbf{Speech quality.} The early attempts at speech quality metrics (eg. PESQ \cite{pesq}, POLQA \cite{polqa}, ViSQOL \cite{hines2015visqol}) were developed based on human studies. However, the methods were found to be sensitive to distortions \cite{hines2013robustness, manjunath2009limitations}. Some of the later works tried to improve PESQ and STOI using expensive gradient updates \cite{zhang2018training, fu2019learning}. DPAM \cite{manocha2020differentiable} learns a perceptual metric by learning a model from crowdsourced human judgments asked to answer whether the two recordings are identical. They show their metric better correlates with MOS tests compared to PESQ \cite{pesq}. However, the metric requires a large set of human judgments and can still generalize poorly to new speakers and content. CDPAM \cite{manocha2021cdpam} aims to use a combination of contrastive and multi-dimensional representation learning to separately model two similarities- content and acoustic. Concurrently, SESQA \cite{serra2021sesqa} uses 5 complementary tasks to improve performance. \\
The above speech quality metrics can be used for both speech enhancement and TTS. However, for TTS, metrics like WER, SpeechLMScore \cite{maiti2023speechlmscore}, MOSNet \cite{lo2019mosnet} are prominently used. Voicebox \cite{le2023voicebox} introduces Fréchet Speech Distance (FSD) by adapting Fréchet distance using self-supervised wav2vec 2.0 features. \cite{alharthi2023evaluating} propose an evaluation technique involving the training of an ASR model on synthetic speech and assessing its performance on real speech. The gold standard evaluation is subjective metrics based on MOS along the direction of naturalness and intelligibility. \\
\textbf{Sound quality.} TTA generation focuses on synthesizing general audio based on text descriptions. The metrics used for objective evaluation include Frechet distance (FD), Frechet Audio Distance (FAD), Inception Score (IS), and Kullback–Leibler (KL) divergence. All the above metrics require the computation of audio embedding, specifically VGGish \cite{cnn_architectures_audioset} for FAD and PANN \cite{pann} for others. For subjective evaluation, two aspects are evaluated. The users are asked to rate the generated samples for their (a) Overall quality (OVL) and (b) relevance to input (REL) on a scale of 1 to 100 or 1 to 5. \\
\textbf{Music quality}. TTM generation focuses on synthesizing music based on text descriptions. The objective and subjective metrics used are the same as TTA generation. MusicLM \cite{musiclm} also uses MuLan \cite{huang2022mulan} to compute the file-wise similarity between text and audio embeddings. For subjective evaluation, MusicLM uses an A-vs-B human rating task, to check the adherence of generated samples to the text descriptions. The users are required to choose between two samples by selecting one of the five answers: strong or weak preference for A or B, and no preference. \\
\textbf{Audio-Text metrics} The existing audio-text metric in literature, CLAP score \cite{liu2023audioldm}, measures the similarity between the caption and the generated audio. The metric measures the relevance between audio and text. 

\section{Effect of distortions} \label{appendix: effect of distortions}
\begin{figure*}
    \centering
    \subfigure{\includegraphics[width=0.19\textwidth]{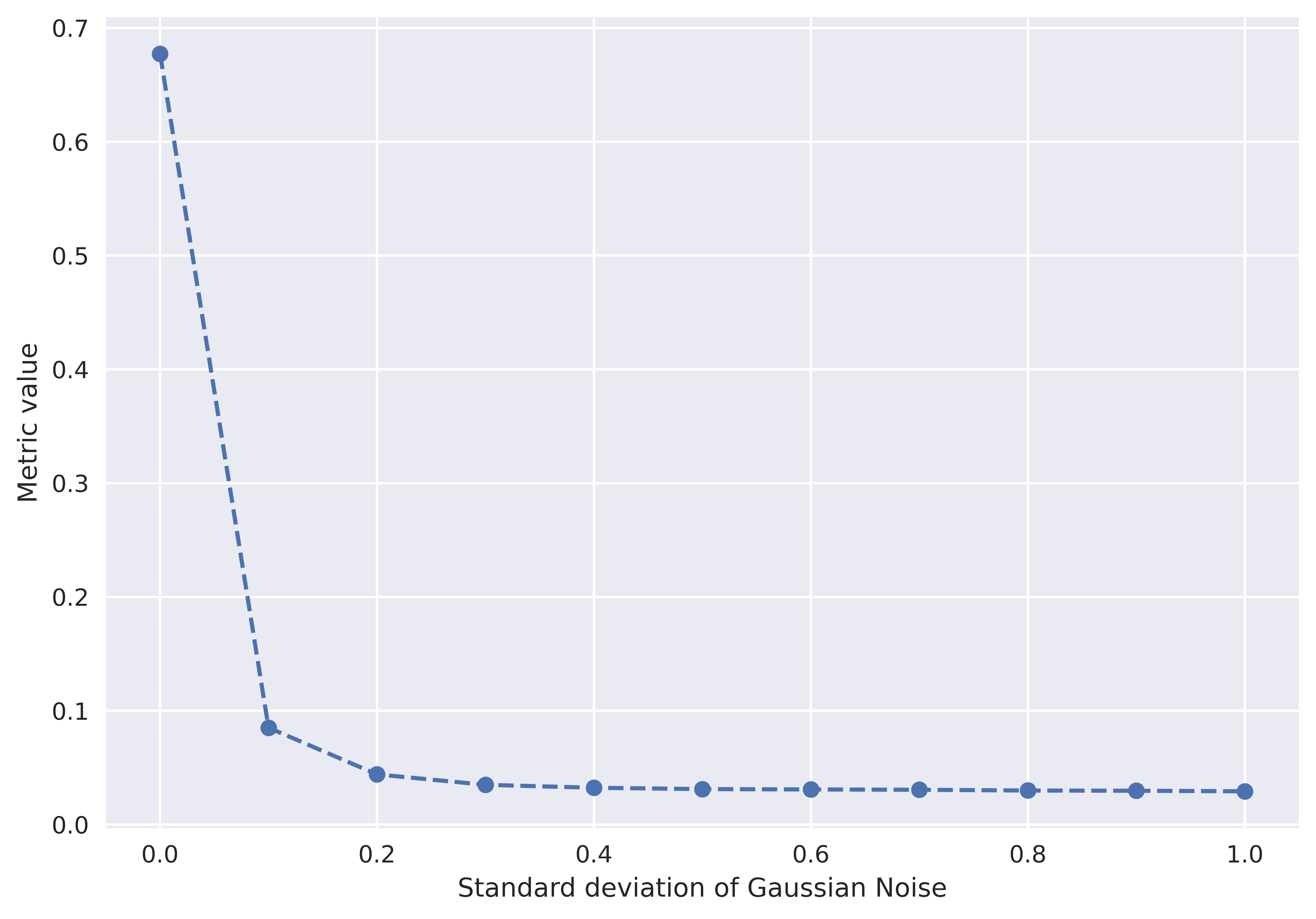}} 
    \subfigure{\includegraphics[width=0.19\textwidth]{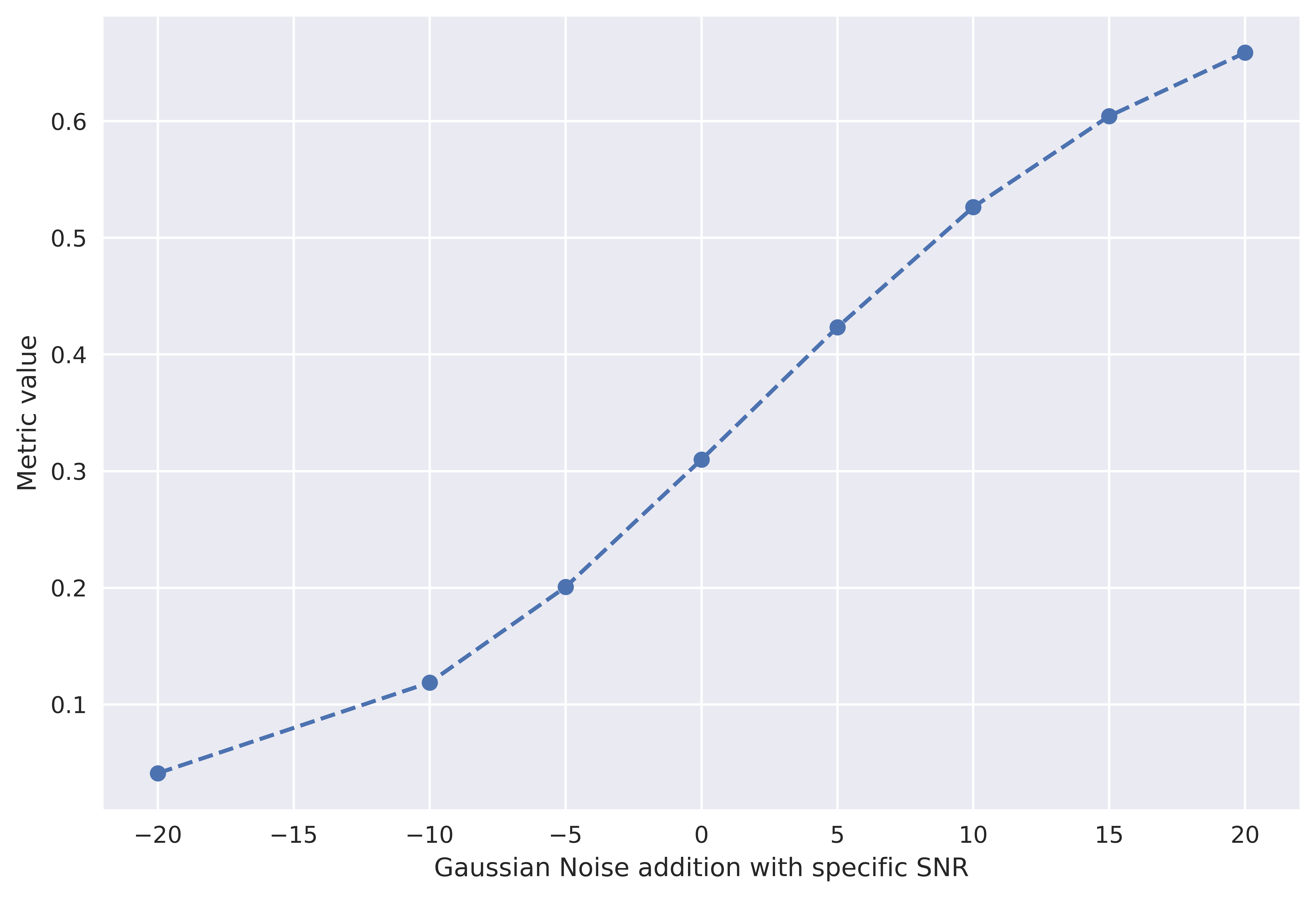}}
    \subfigure{\includegraphics[width=0.19\textwidth]{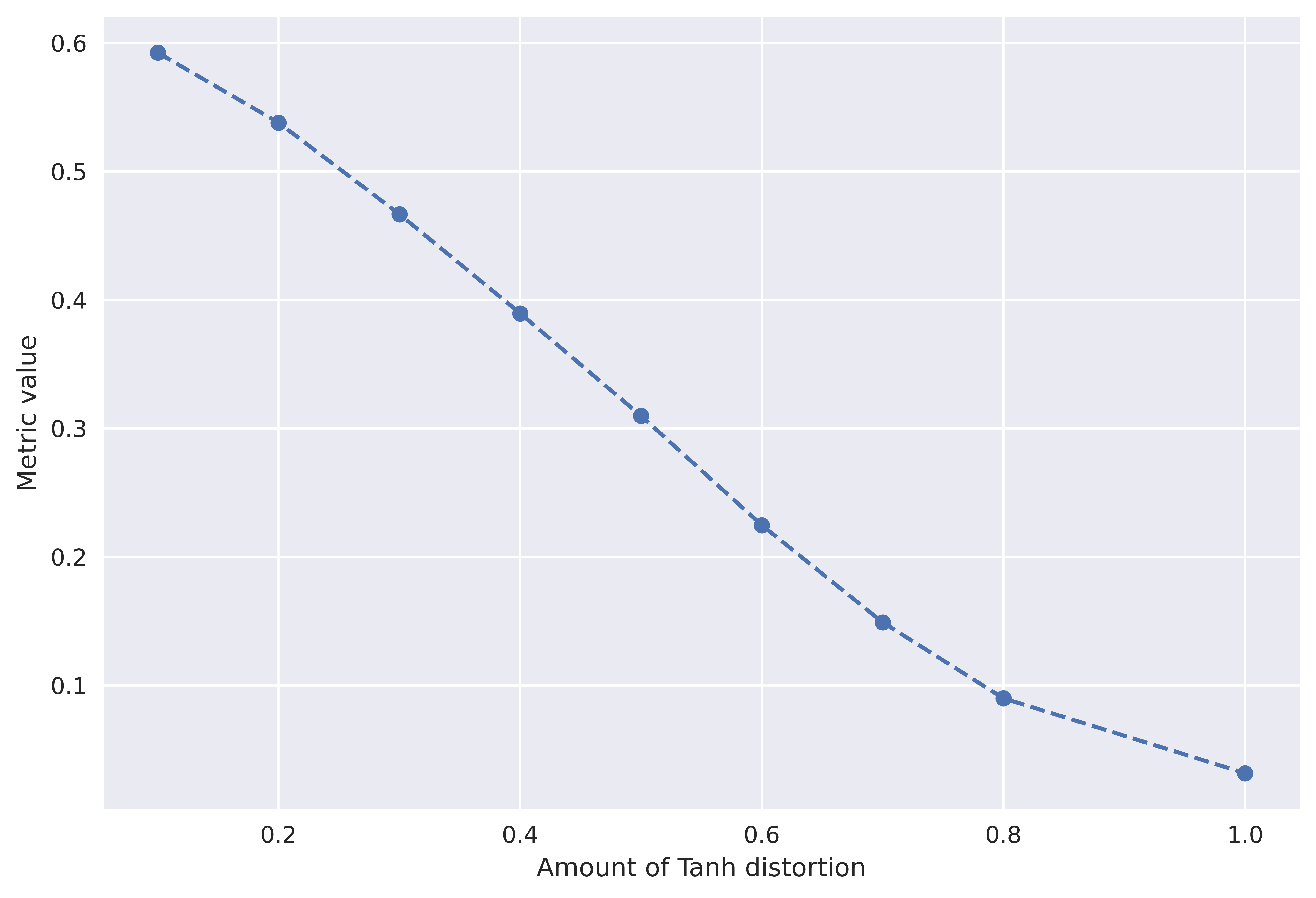}}
    \subfigure
    {\includegraphics[width=0.19\textwidth]{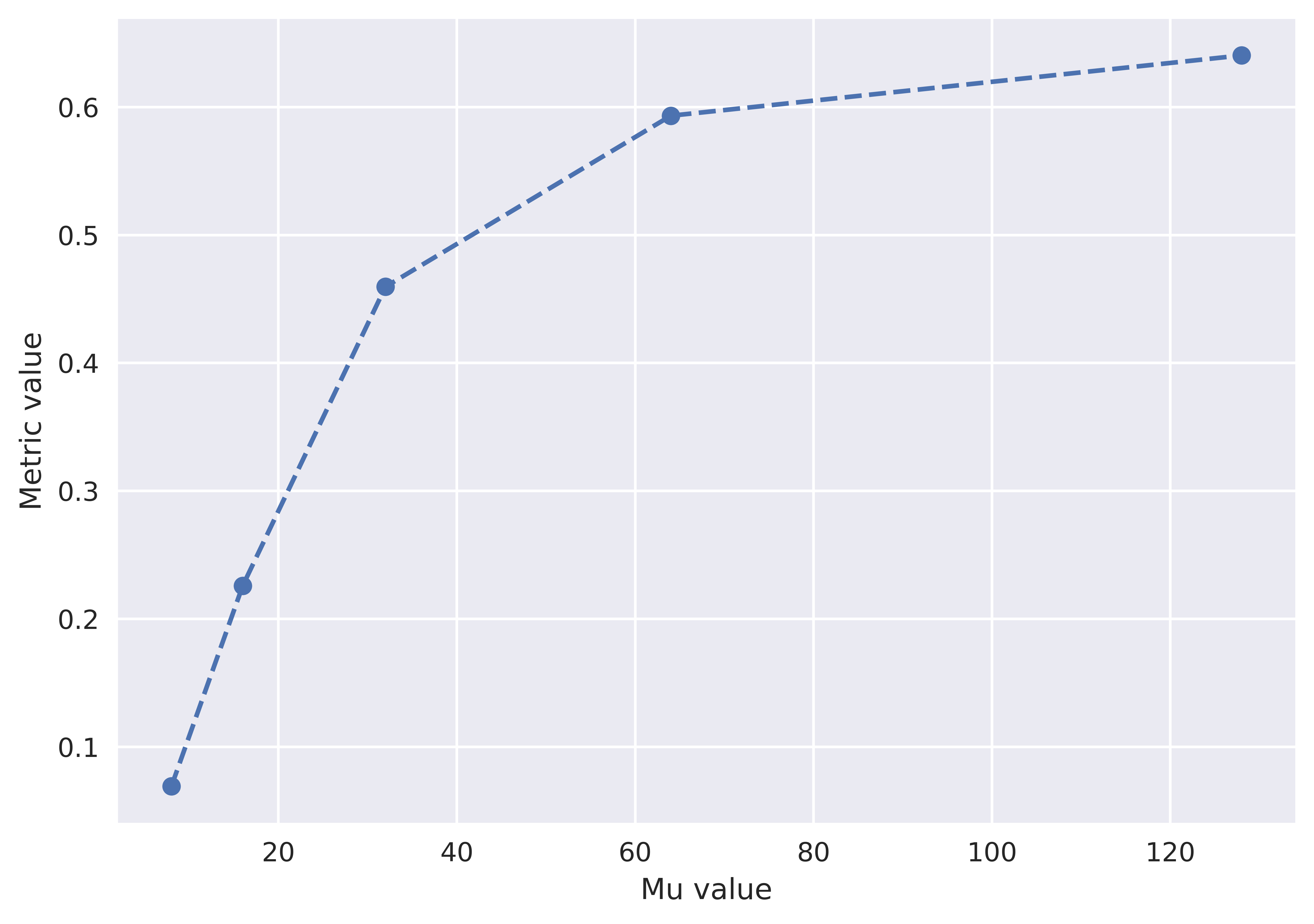}}
    \subfigure{\includegraphics[width=0.19\textwidth]{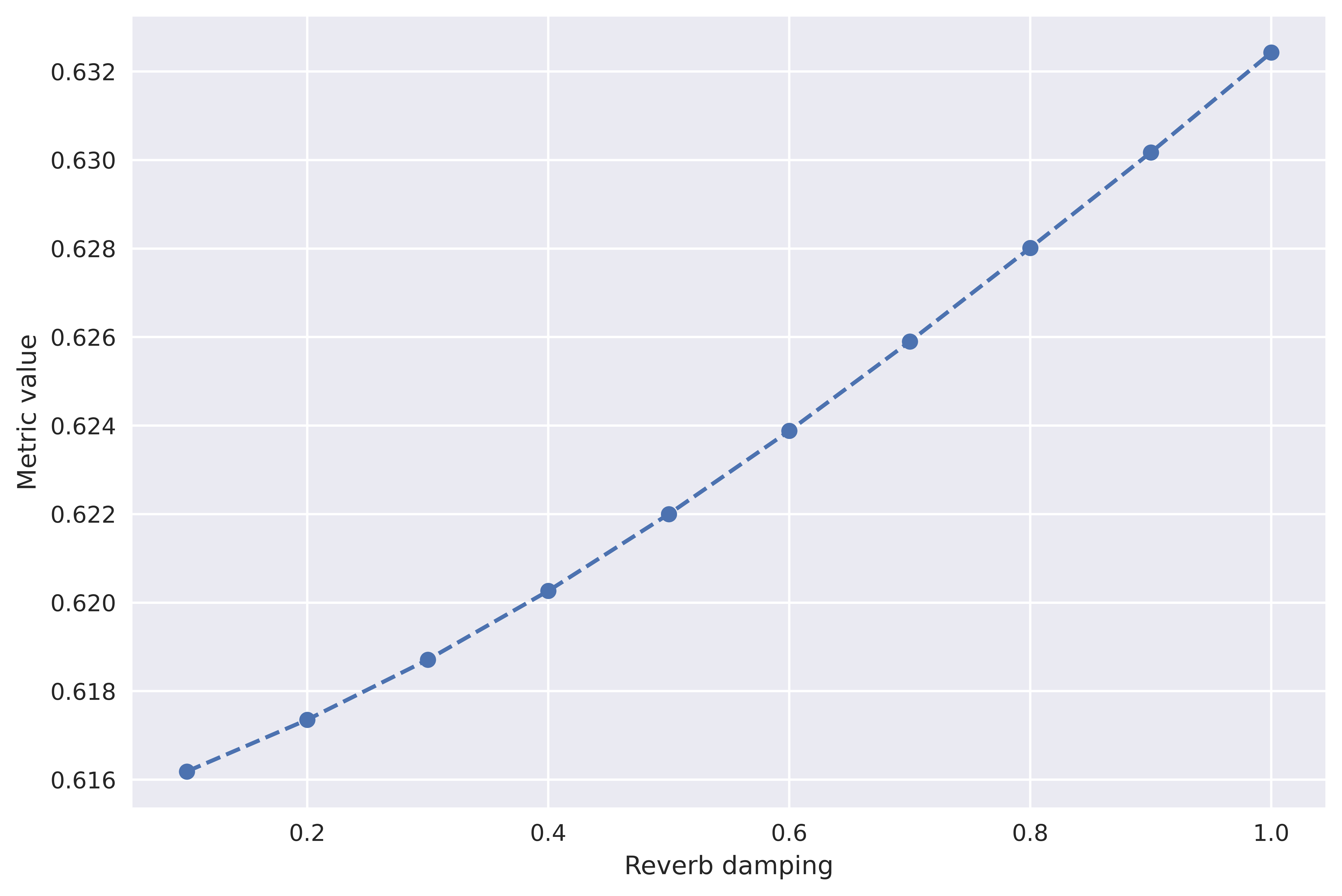}}
    {\includegraphics[width=0.19\textwidth]{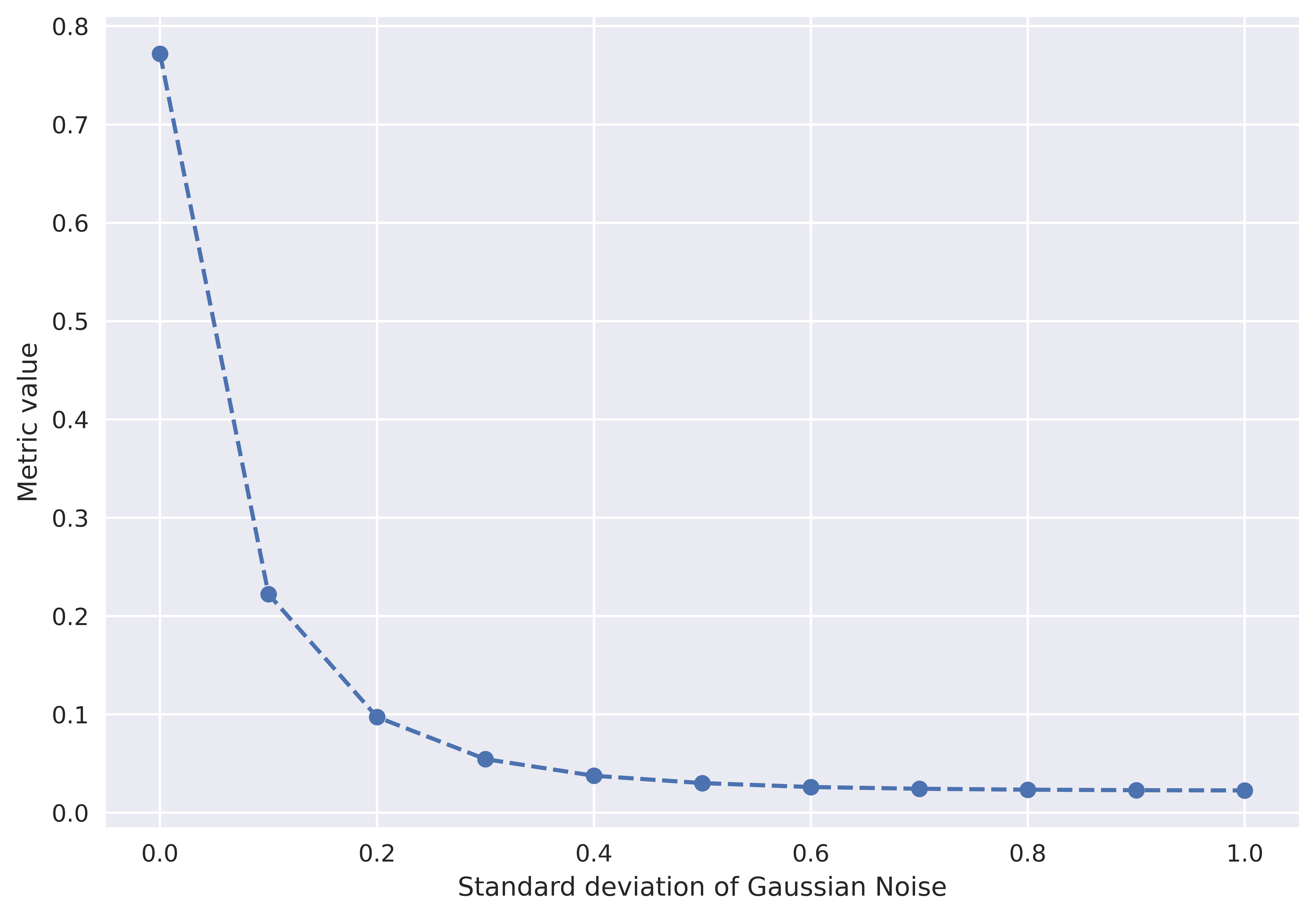}} 
    \subfigure{\includegraphics[width=0.19\textwidth]{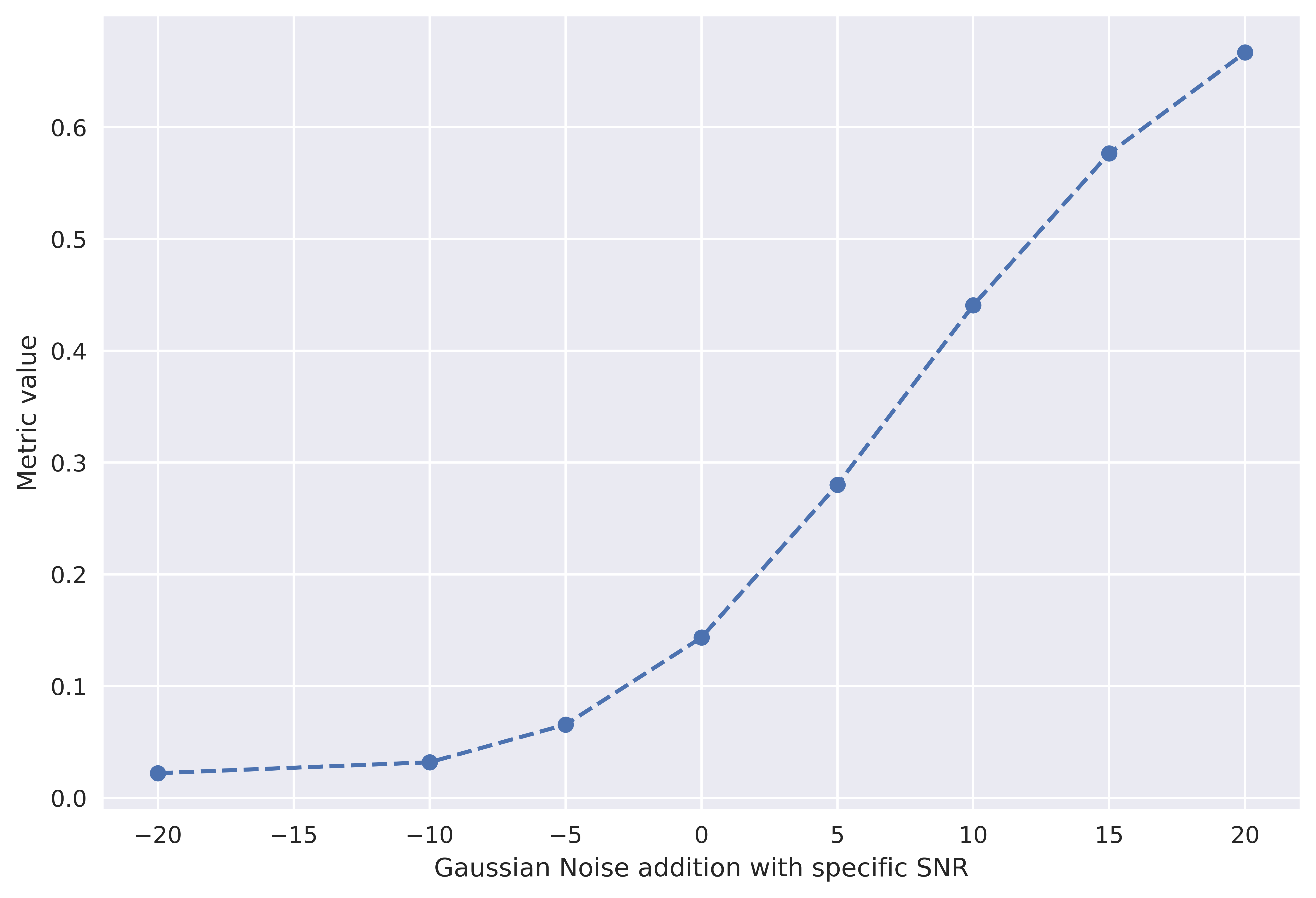}} 
    \subfigure{\includegraphics[width=0.19\textwidth]{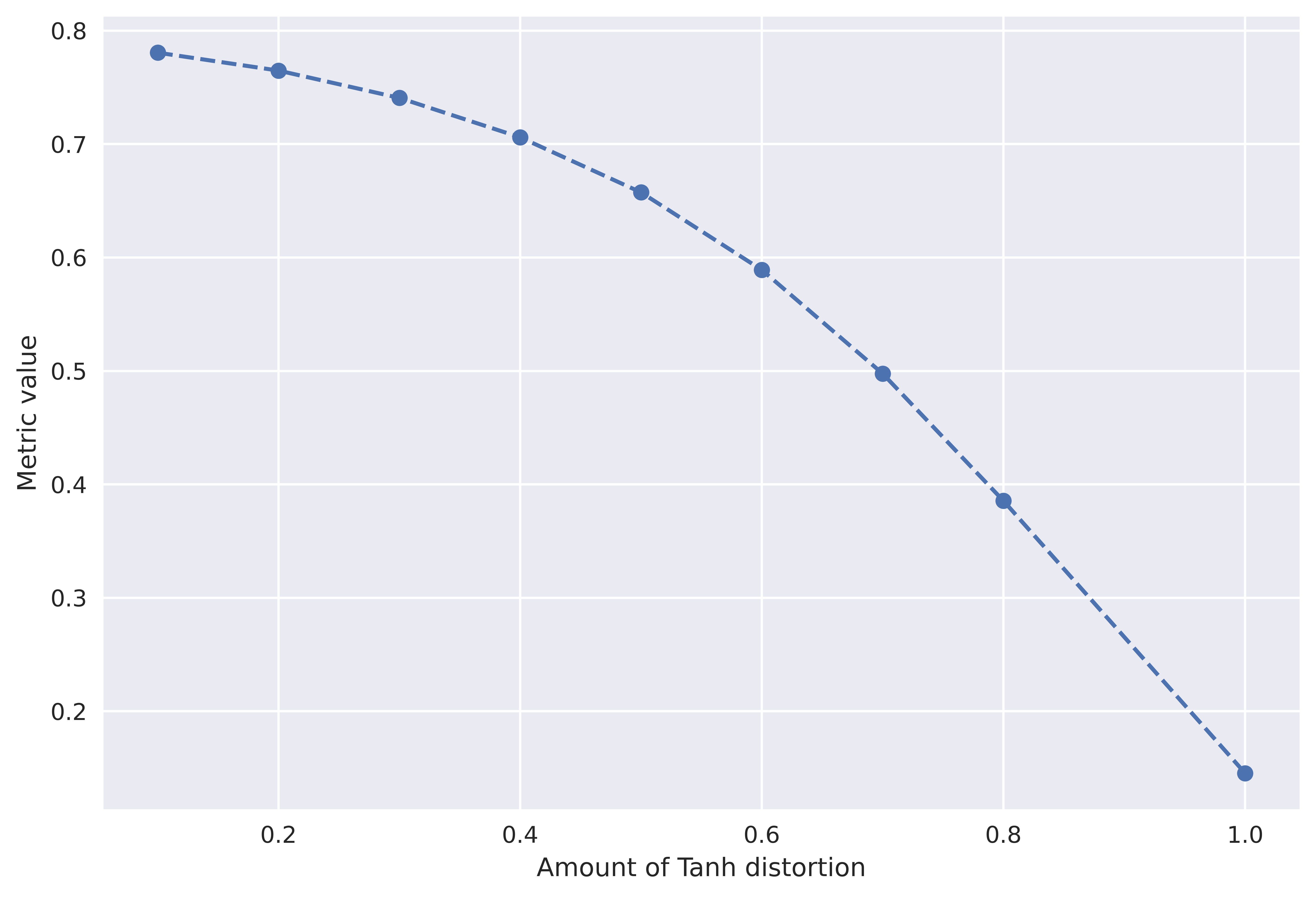}}
    \subfigure
    {\includegraphics[width=0.19\textwidth]{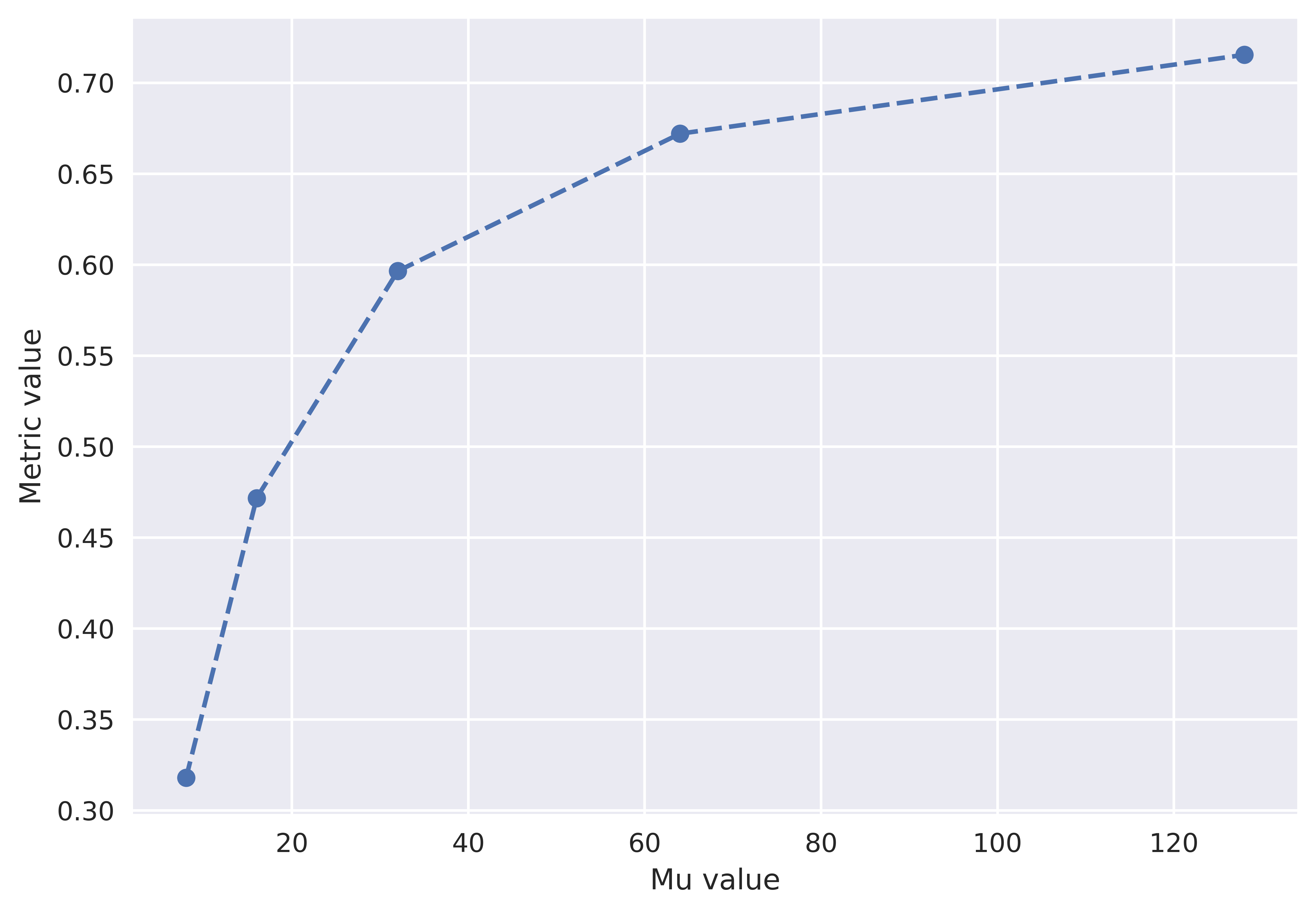}}
    \subfigure{\includegraphics[width=0.19\textwidth]{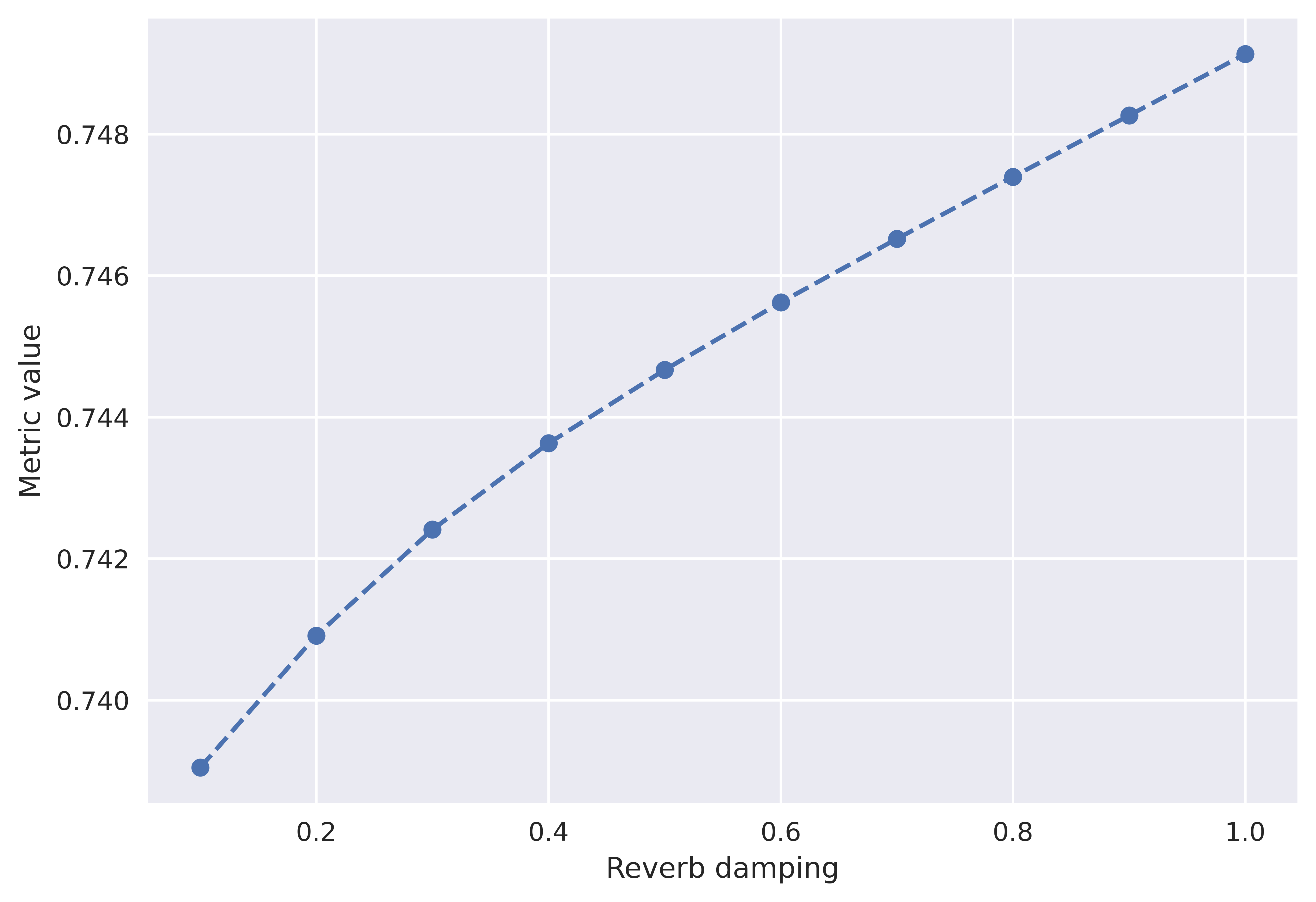}}
    {\includegraphics[width=0.19\textwidth]{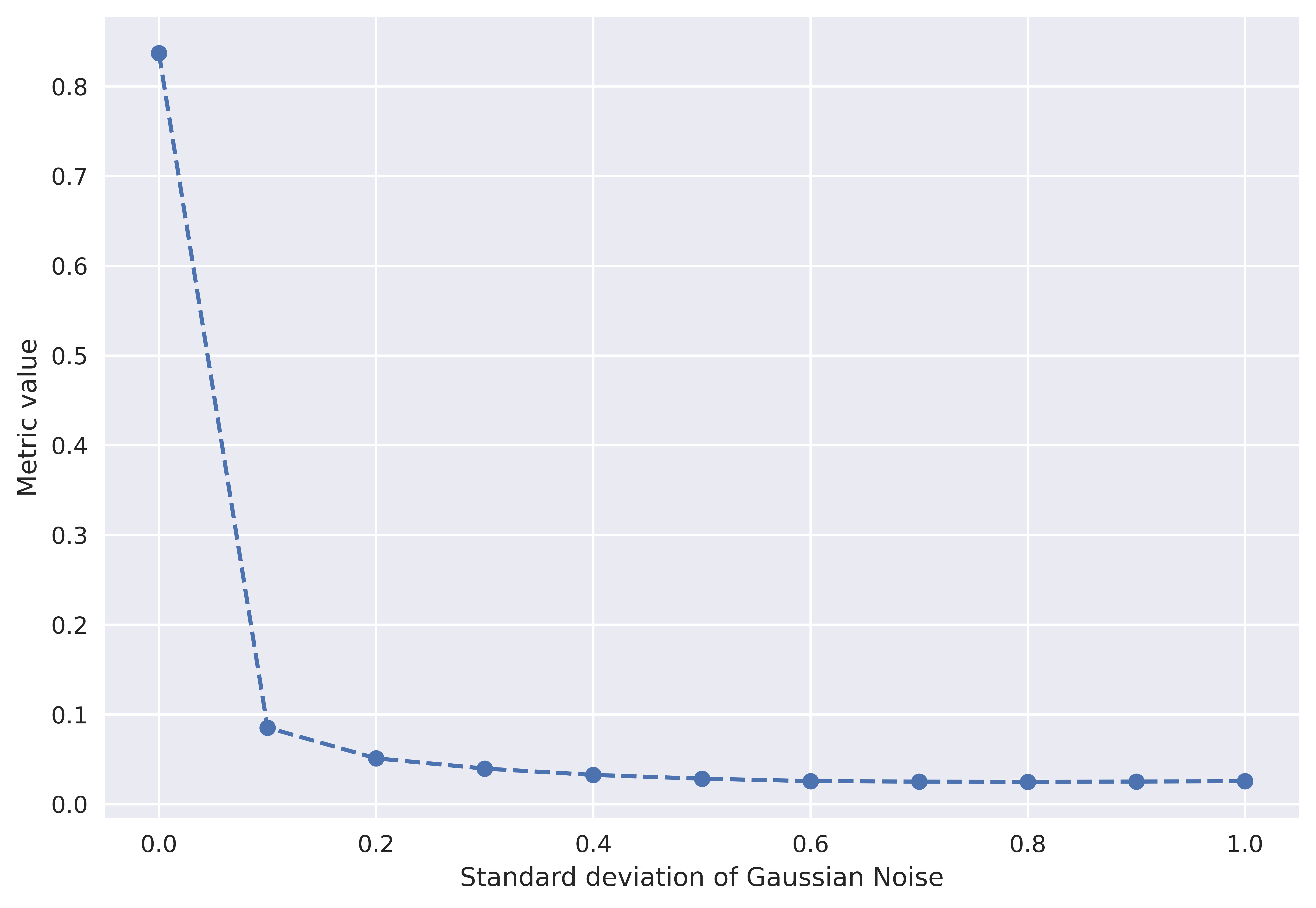}} 
    \subfigure{\includegraphics[width=0.19\textwidth]{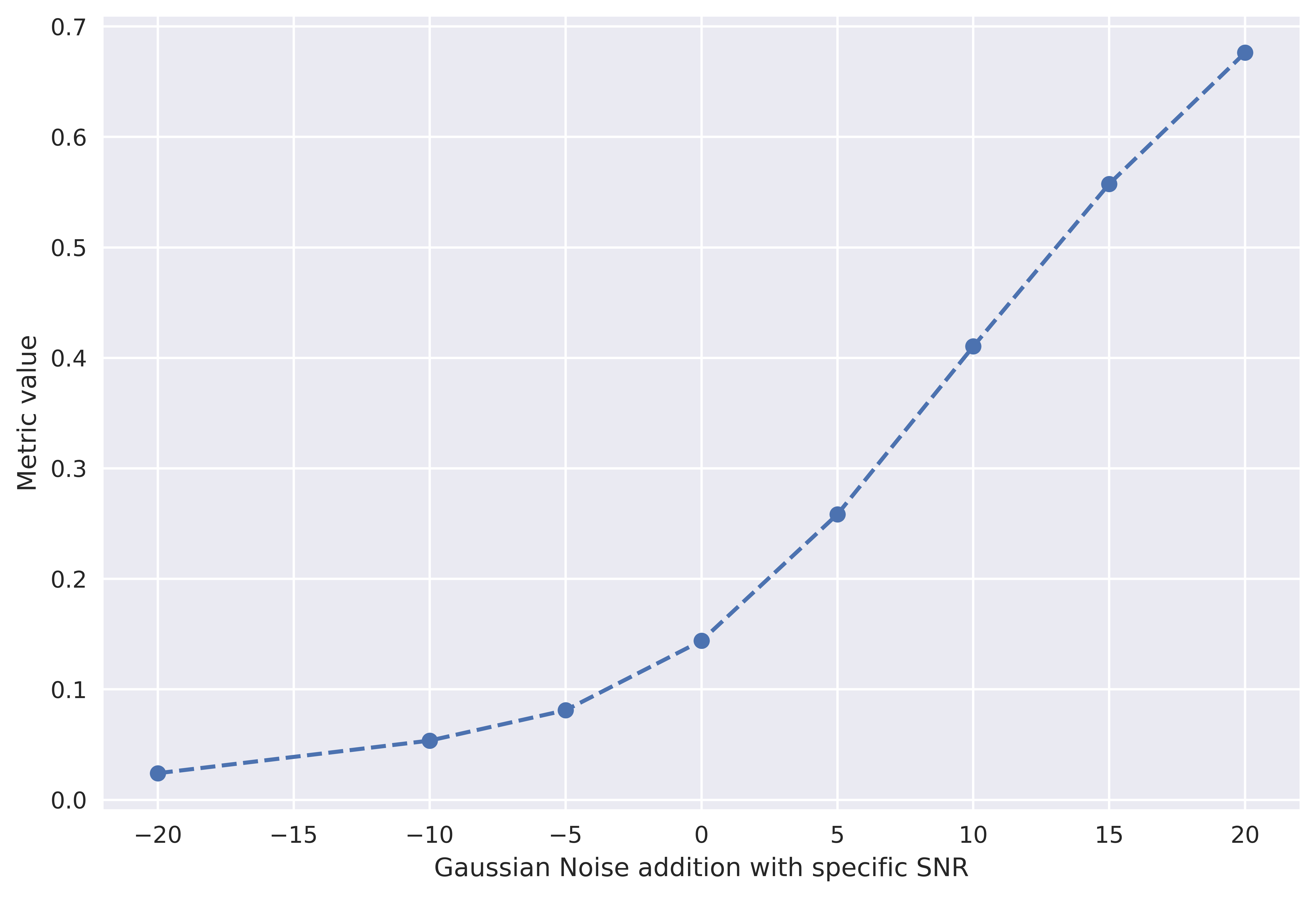}} 
    \subfigure{\includegraphics[width=0.19\textwidth]{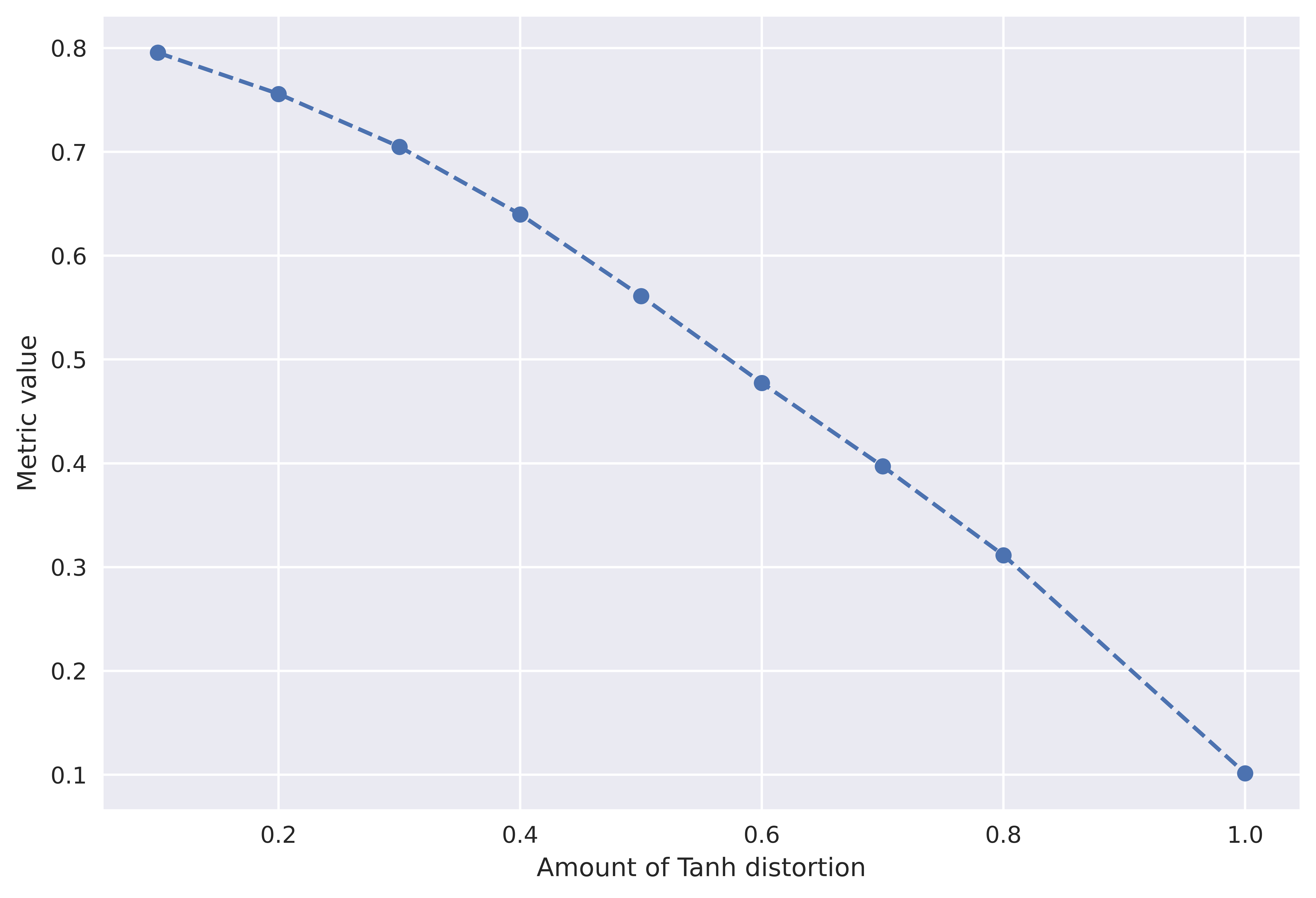}}
    \subfigure
    {\includegraphics[width=0.19\textwidth]{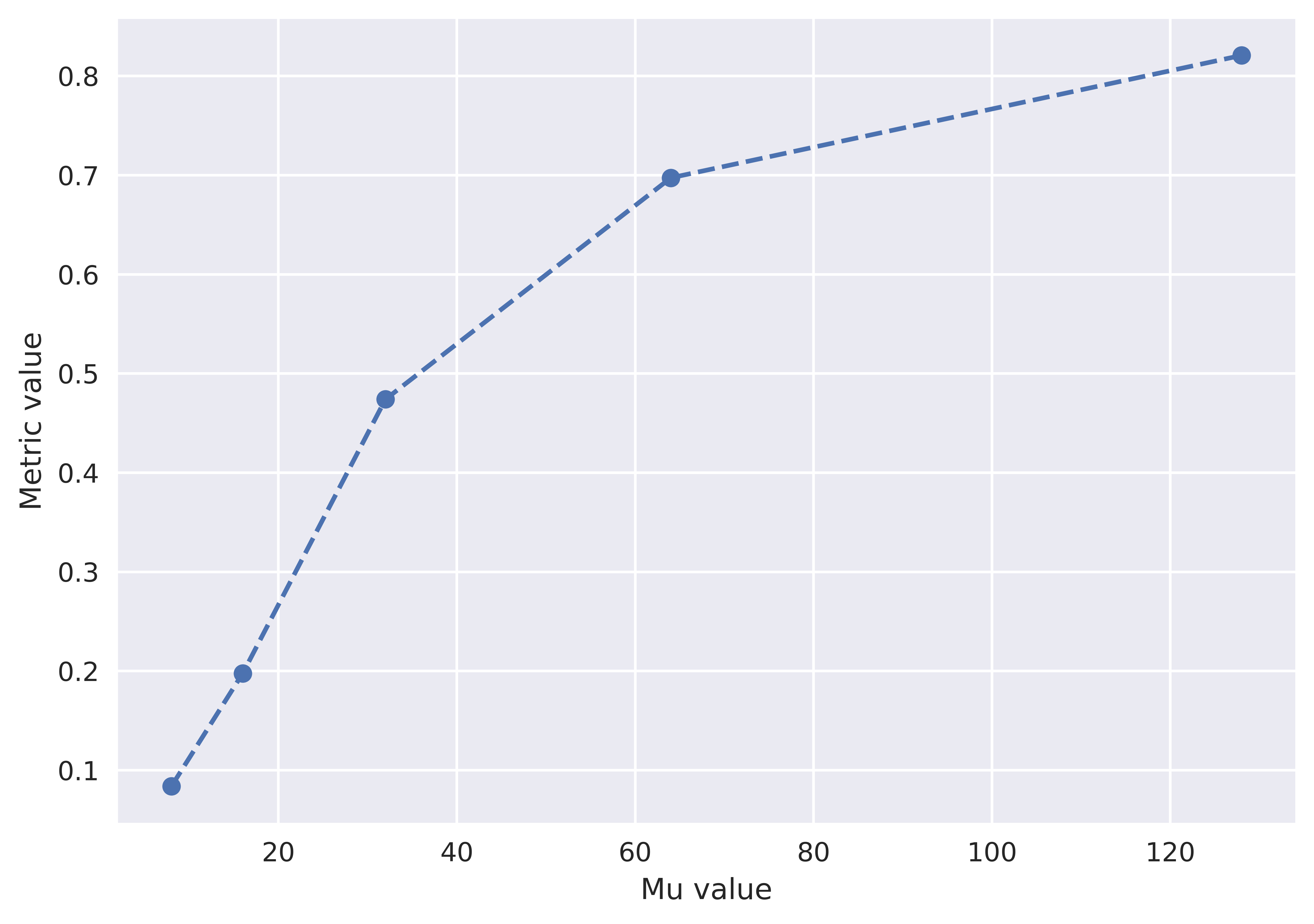}}
    \subfigure{\includegraphics[width=0.19\textwidth]{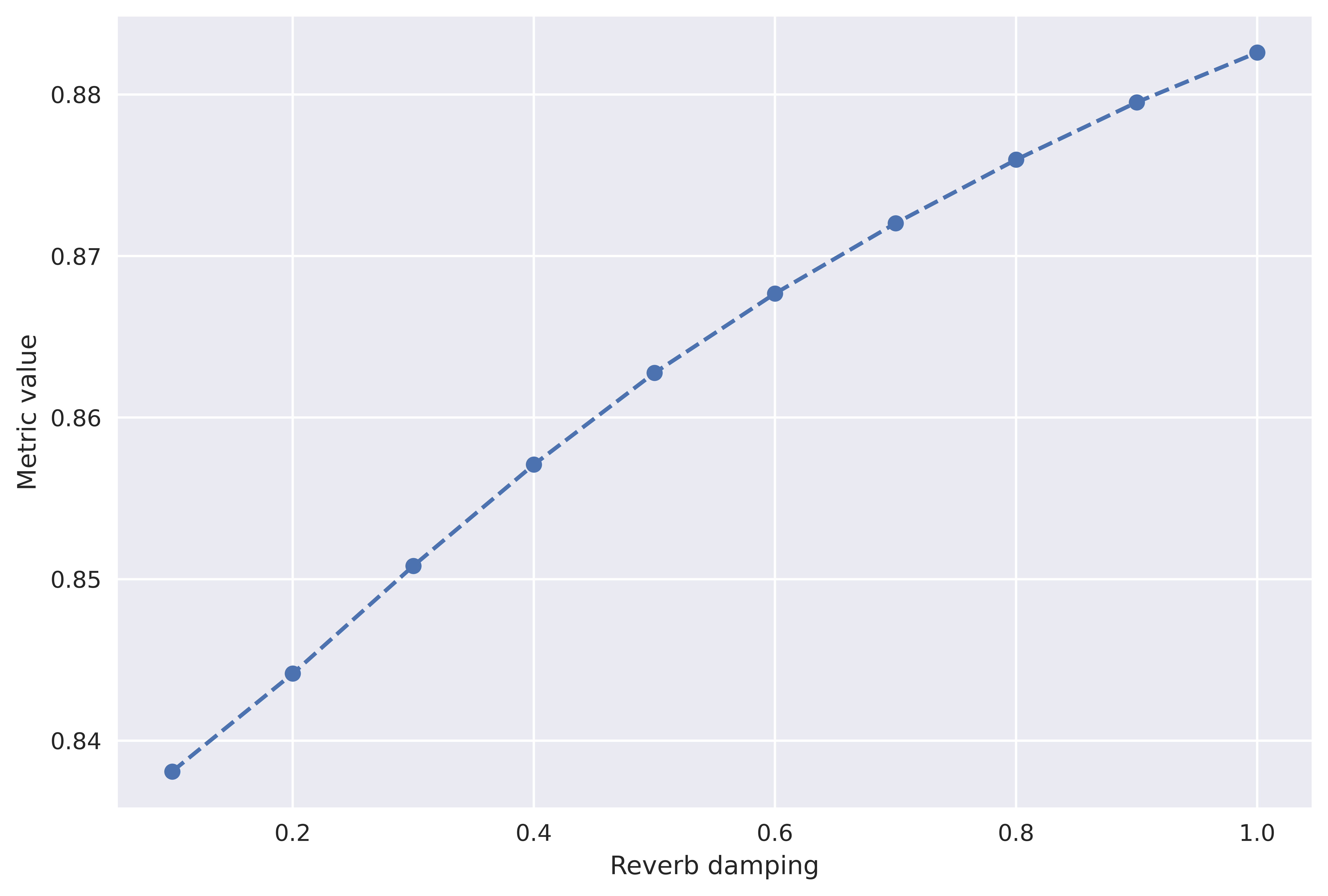}}
    \caption{Effect of (a) Gaussian Noise (b) Gaussian Noise with SNR (c) Tanh distortion (d) Mu-Law compression (e) Reverb. on \metricname value. The first row uses AudioCaps sourced from YouTube, the second row uses MusicCaps sourced from YouTube, and the third row uses LibriTTS clean set. The Figure \ref{fig:distortions_main} shows effect of distortion for professionally recorded audio}
    \label{fig: appendix distort}
\end{figure*}

This section \ref{sec:result: effect of distortions} shows results on a professionally recorded audio pack. In this section, we vary the source data and check degradation in \metricname score. The source data considered is Professionally recorded audio, AudioCaps (Sound events, YouTube sourced), MusiCaps (Music, YouTube sourced), and LibriTTS (speech, audiobooks). The four types of distortions used are (1) Gaussian noise with increasing standard deviation (2) Gaussian Noise addition with particular SNR (3) Tanh distortion (4) Mu Law compression (5) Reverb. Lastly, we also add Reverb, which by the definition in Section \ref{sec: audio quality} is not considered as an artifact or distortion. Figure \ref{fig: appendix distort} shows the effect of distortions on \metricname score across different source distributions. We see the \metricname score degrading with the addition of noise except for Reverb where it remains constant.

\section{Text-to-Audio generation} \label{appendix: text-to-audio generation}

\subsection{Text-to-Audio models} \label{appendix: text-to-audio models arch}
For TTA generation, we use publicly available variants of AudioLDM \cite{liu2023audioldm}, AudioLDM2 \cite{audioldm2}, AudioGen \cite{kreuk2022audiogen} and MelDiffusion. \\
\textbf{AudioLDM.} The model \cite{liu2023audioldm} is based on latent diffusion models (LDMs) \cite{rombach2022high}. The latent space is obtained by applying a variational autoencoder (VAE) to the mel-spectrograms of audio clips. The LDMs use UNet conditioned on CLAP text embeddings. During training, the LDMs learn to reconstruct the audio embeddings from Gaussian noise, while being guided by the text embeddings. During sampling, the LDMs generate audio embeddings from the text embeddings and then decode them into waveforms using the VAE followed by a HiFi-GAN \cite{kong2020hifi} vocoder. Our experiments use the model versions hosted on huggingface with 100 denoising steps to generate audio. \\
\textbf{AudioLDM2.} The model consists of three main components: a text encoder, a GPT-2 decoder, and a latent diffusion model. The text encoder uses two pre-trained models, CLAP and Flan-T5, to obtain text embeddings that capture both the alignment and the semantics of the text. Then GPT-2 generates a sequence of new embedding vectors, called the language of audio (LOA), based on the text embeddings. The latent diffusion model de-noises a random latent vector into an audio waveform, conditioned on the LOA and the Flan-T5 text embeddings. The model is trained with self-supervised pre-training and fine-tuning on different audio domains. Our experiments use the model versions hosted on huggingface with 100 denoising steps to generate audio. \\
\textbf{AudioGen.} The model is a Transformer decoder operating over a residual vector quantized representation (RVQ) of the audio signal. The model generates audio from text by using textual features as conditioning signal. Our experiments use the model versions hosted on huggingface \cite{kreuk2022audiogen}. The model uses Encodec \cite{defossez2022high} to obtain the RVQ from audio, T5 \cite{2020t5} to obtain textual features and is trained using the delay-pattern technique \cite{musicgen} to model the RVQ. \\
\textbf{MelDiffusion.} The model is based on using the diffusion model on spectrograms instead of latent space. The text encoder used is T5-large \cite{raffel2020exploring} and the diffusion model is DDIM based on progressive distillation \cite{salimans2022progressive}. The base UNet is inserted with additional self-attention layers to produce coherent 30-second or more audio while training on only 5-second audio. For inference, we use 100 denoising steps to generate audio. 

\subsection{Human Listening experiment} \label{appendix: t2a human listening}
We evaluated text-to-audio generation using Amazon Mechanical Turk (MTurk). In this evaluation, participants were asked to rate the quality of the audio and its relevance to the provided description. The ratings were given on a  Likert scale from 1 (poor quality or minimal relevance) to 5 (excellent quality or perfect match with the description). Detailed instructions given to participants are outlined in Table \ref{tab:instruction}, and the specific questions posed, along with their response options, are listed in Table \ref{tab:questions}. For this test, we chose 100 random samples from the AudioCaps dataset. We then generated audio for these samples using four different models: MelDiffusion, AudioLDM2-l, AudioLDM-l, and AudioGen-m. resulting in 500 samples. Each of these samples was rated by 10 different participants, all of whom were located in the United States, resulting in a total of 8,000 scores evaluating both the quality and relevance of the audio. To ensure the quality and reliability of the data, we applied a rigorous filtering process to the responses. If a participant's scores showed a standard deviation of zero for more than five samples, their responses were excluded from the analysis. Also, any responses from participants who took less than 10 seconds to complete their ratings were also excluded. Furthermore, we will release the collected data, both raw and filtered.

\begin{table}[ht!]
\centering
\caption{Guidelines Given to Amazon Mechanical Turk Participants for TTA task.}
\label{tab:instruction}

\begin{tabular}{p{1\linewidth}}
\hline\hline
\\
\textbf{Task Instructions} \\
Your task is to evaluate the quality and text relevance of audio clips. These clips include various sounds and speech such as dog barking and rain. You will first rate the sound quality, and then assess its relevance to the text description. \\
 \\
\textbf{Definition of quality in this test: } \\
In this evaluation, 'quality' refers to the fidelity of the generated audio in replicating real-life sounds. Our focus is on assessing a text-to-audio generation system, which converts textual descriptions into corresponding audio outputs. The audio output may include real-world noises, such as ambulance sirens, dog barking, and screaming. The primary goal is to assess the realism of these sounds in the audio.\\
\\

\textbf{Important Note:} \\

Please be aware that during this audio quality test, you may encounter segments where speech is present. It's normal and expected that the speech might not be intelligible. This is not a concern for this specific test. Your main focus should be on evaluating the overall audio quality, not the intelligibility of the words spoken.\\
\\

\textbf{Warning: } \\
Please be advised that during this audio test, some segments may feature very loud sounds. We recommend adjusting your volume to a comfortable level before beginning the test and being prepared to adjust it as needed during the test. Your safety and comfort are important to us. If at any point you find the audio uncomfortably loud, please feel free to lower the volume or pause the test to readjust your settings.\\
\\

\hline \hline
\end{tabular}
\end{table} 

\begin{table}[ht!]
\centering
\caption{Questions and Response Options Presented to MTurk Participants for TTA task.}
\label{tab:questions}

\begin{tabular}{p{1\linewidth}}
\hline\hline
\\
\textbf{Please listen carefully to the following audio then answer the two questions below.} 
\\ \\

\textbf{How good does the audio sound to you in terms of quality and realism?}
\\ \\

\textbf{1 (Poor)} The audio quality is very low, making it hard to discern the intended sounds.\\
\textbf{2 (Fair)} Audio quality is below average, but the intended sounds are somewhat recognizable.\\
\textbf{3 (Good)} The audio has decent quality with clear and recognizable sounds.\\
\textbf{4 (Very Good)} Audio quality is high, closely resembling real-world audio with minimal distortion.\\
\textbf{5 (Excellent)} The audio quality is highly realistic with perfect fidelity.
  
\\
\\

\textbf{How well does this audio match with the provided description?}\\

\textbf{Description:} audio description.\\
\\

\textbf{1 (Poor)} Audio has minimal or no relevance to the text.\\
  
\textbf{2 (Fair)} Audio shows limited relevance to the text.\\
\textbf{3 (Good)} Audio is adequately relevant to the text.\\
  
\textbf{4 (Very Good)} Audio is highly relevant to the text.\\
  
\textbf{5 (Excellent)} Audio perfectly matches the text.\\
\\

\hline \hline

\end{tabular}
\end{table}

\section{Text-to-Music generation} \label{appendix: text-to-music generation}

\subsection{Text-to-Music models} 
For TTM generation, we use AudioLDM2-m \cite{audioldm2}, MusicLDM \cite{musiclm}, and MusicGen \cite{musicgen}.\\
\textbf{AudioLDM2-m.} The architecture is from AudioLDM2 (Section \ref{appendix: text-to-audio models arch}) but trained on music. \\
\textbf{MusicLDM.} The model adapts Stable Diffusion and AudioLDM architectures to the music domain. For this, the CLAP and vocoder components are retrained along with the introduction of a beat-tracking model and different mixup strategies. The mixup strategies encourage the model to generate music that is diverse yet grounded in the requested style. \\
\textbf{MusicGen.} The model is similar in architecture and training objective as AudioGen, described in \ref{appendix: text-to-audio models arch}, but trained on music rather than audio events. Our experiments use the model versions hosted on huggingface using the default provided configuration. Namely, each model generates music with a 32kHz sampling rate, discretized using an Encodec tokenizer with 4 codebooks where each token is sampled at 50 Hz. The token generation uses \texttt{top-k} sampling where $k=250$. \\

\subsection{Human Listening experiment} \label{appendix: t2m human listening}
To evaluate the effectiveness of \metricname in TTM generation, we conducted a human evaluation test using MTurk. In this test, participants were asked to rate the quality of music generated and its relevance to a given description. These ratings were based on a Likert scale ranging from 1 (poor quality or minimal relevance) to 5 (excellent quality or perfect match with the description), as detailed in Table \ref{tab:music_questions} and \ref{tab:instruction_ttm}. For this purpose, we selected 100 random samples from MusicCaps dataset. For each sample, we generated music based on a text description using four different models: AudioLDM2-m, MusicLDM, MusicGen-l, MusicGen-mel, and the original MusicCaps model resulting in a total of 500 audio samples. To ensure a comprehensive evaluation, each sample was rated by 10 different participants, all of whom were located in the United States, culminating in 8,000 individual scores assessing both quality and relevance. In order to maintain the integrity of our data, we applied a filtering process similar to the one used in our TTA generation test. We excluded any participant whose ratings showed no variation (a standard deviation of zero) for more than five samples, or who completed the rating in less than 10 seconds. We will release both the raw and filtered datasets for human evaluation. This will allow for further analysis and transparency in our findings.

\begin{table}[ht!]
\centering
\caption{Guidelines Given to MTurk Participants for TTM task.}
\label{tab:instruction_ttm}

\begin{tabular}{p{1\linewidth}}
\hline\hline
\\
\textbf{Task Instructions} \\

Your task is to rate the overall quality of the music and the relevance of the music with the text description. Listen to each clip and first evaluate its sound quality. Then, assess how well the music matches the provided description.\\

\textbf{Important Note:} \\

 During this music evaluation test, you might find descriptions mentioning a singer or vocals. However, please note that the actual audio may consist only of instrumental music without any singing. This discrepancy is normal and expected for this test. Even if the description refers to singing, your focus should be on assessing the music's quality and how well the instrumental audio aligns with the overall theme of the description, irrespective of the presence of singing.\\
\\

\hline \hline
\end{tabular}
\end{table}

\begin{table}[ht!]
\centering
\caption{Questions and Response Options Presented to MTurk Participants for TTM task.}
\label{tab:music_questions}

\begin{tabular}{p{1\linewidth}}
\hline\hline
\\

\textbf{Please listen carefully to the following audio then answer the two questions below.} 
\\ \\

\textbf{How good is the quality of the music?}
\\ \\

\textbf{1 (Poor)} The music quality is very low, with poor clarity and composition.\\
\textbf{2 (Fair)} Music quality is below average, with some elements of composition recognizable.\\
\textbf{3 (Good)} The music has decent quality with clear composition and a pleasant listening experience.\\
\textbf{4 (Very Good)}  Music quality is high, offering a rich and engaging listening experience.\\
\textbf{5 (Excellent)} The music quality is outstanding with excellent clarity, composition, and overall appeal.
  
\\
\\

\textbf{How well does this music match with the provided description?}\\

\textbf{Description:} audio description.\\
\\

\textbf{1 (Poor)} Music has minimal or no relevance to the description.\\
  
\textbf{2 (Fair)} Music shows limited relevance to the description.\\
\textbf{3 (Good)} Music is adequately relevant to the description.\\
  
\textbf{4 (Very Good)} Music is highly relevant to the description.\\
  
\textbf{5 (Excellent)} Music perfectly matches the description.\\
\\

\hline \hline

\end{tabular}
\end{table}

\section{Noise suppression} \label{appendix: noise suppression}
\subsection{Problem description}
DNS aims at enhancing speech for voice communication by removing unwanted noise from a recording. 
However, DNS typically introduces its own processing artifacts and distortions that may degrade the desired speech signal or cause unpleasant artifacts in the  background noise that is not suppressed. Therefore, the performance of a DNS model in terms of perceptual quality depends on a variety of factors. To measure the quality of DNS systems, a subjective listening test can be performed where human judges assign ratings to the model output, typically from 1 (worst) to 5 (best). The Mean Opinion Score (MOS) for an output sample is obtained by averaging the human ratings. As an alternative to costly subjective testing, machine-learning models can be trained on DNS output samples and their corresponding MOS labels to perform blind DNS MOS estimation. Various DNS models or model variations can be compared in terms of their average subjective or estimated MOS. In Section \ref{sec:DNS} the performance of \metricname for ranking various DNS models is compared to a state-of-the-art DNS MOS estimation model. The comparison is performed on the blind test set of the ICASSP 2021 DNS challenge processed by over 20 different DNS models. The state-of-the-art DNS-MOS estimator and \metricname are compared in terms of the Spearman's Rank Correlation Coefficient (SRCC) computed using the MOS averaged for each model. The authors of DNS-MOS found this to be a robust metric for evaluating the performance of a MOS estimator for comparing different DNS models. 

\subsection{Prompt averaging}
Given the complex and multifaceted nature of the perceptual quality of DNS output samples, we experiment with two simple prompt averaging schemes that aim at a broader and more robust quality estimation: \metricnameAvgSim and \metricnameAvg.
The underlying hypothesis is that averaging over multiple quality-related prompts may yield a less noisy and perceptually broader similarity metric than the two primary prompts ($\mathrm{h}1$ and $\mathrm{b}1$ below) that focus specifically on the presence or absence of noise and artifacts.
To this end, we introduce two additional prompts directly querying sound \emph{quality}:
\begin{itemize}
    \item \textbf{$\mathrm{h}1$}: ``the sound is clear and clean''
    \item \textbf{$\mathrm{b}1$}: ``the sound is noisy and with artifacts''
    \item \textbf{$\mathrm{h}2$}: ``the sound quality is good''
    \item \textbf{$\mathrm{b}2$}: ``the sound quality is bad''
\end{itemize}
To compute \metricnameAvgSim, we average the dot products before taking the softmax:
\begin{equation}
    z_{\mathrm{h},\mathrm{avg}} = \frac{1}{K}\sum_{i=1}^{K}u_{\mathrm{h}i}\cdot v
\end{equation}
for $K$ high quality prompts $\mathrm{h}i$. 
$z_{\mathrm{b},\mathrm{avg}}$ is computed analogously using low quality prompts $\mathrm{b}i$. 

\metricnameAvgSim is then given as
\begin{equation}
p_\mathrm{h} = \frac{e^{z_{\mathrm{h},\mathrm{avg}}}}{\sum_{j=1}^2 e^{z_{j,\mathrm{avg}}}}.
\label{eq:p_h}
\end{equation}

\metricnameAvg is computed as \metricname averaged over multiple prompt pairs:
\begin{equation}
    p_{\mathrm{h},\mathrm{avg}} = \frac{1}{K}\sum_{i=1}^{K}p_{\mathrm{h}i},
\end{equation}
where $p_{\mathrm{h}i}$ is computed via Eq. \ref{eq:p_h} using a prompt pair $\mathrm{h}i$ and $\mathrm{b}i$. In our preliminary experiments we found the most effective prompt pairs to be $[\mathrm{h}1, \mathrm{b}2]$ and $[\mathrm{h}2, \mathrm{b}1]$, though this finding may not generalize to other tasks or datasets. Note that the proposed simple averaging schemes generalize to arbitrary numbers and combinations of prompts. However, we leave a more thorough investigation of prompting strategies for future work. 
